\documentclass[preprint]{aastex63}
\DeclareUnicodeCharacter{FFFD}{}
\newcommand{\R}[1]{{\color{red}{#1}}}
\submitjournal{ApJL}
\shorttitle{Energy and Mass transport associated with jet flows}
\shortauthors{Ni et al.}

\begin{document}

\title{Energy and mass transport associated with impulsive spicular flows in solar coronal holes}
\correspondingauthor{Lei Ni}
\email{leini@ynao.ac.cn}

\correspondingauthor{Jun Lin}
\email{jlin@ynao.ac.cn}

\author[0000-0001-6366-7724]{Lei Ni}
\affiliation{Yunnan Observatories, 
Chinese Academy of Sciences, 
Kunming, 
Yunnan 650216, P.R.China}
\affiliation{University of Chinese Academy of Sciences, 
Beijing 100049,
P.R.China}
\affiliation{Yunnan Key Laboratory of Solar Physics and Space Science, Kunming 650216, P.R.China}

\author{Jun Lin}
\affiliation{Yunnan Observatories, 
Chinese Academy of Sciences, 
Kunming, 
Yunnan 650216, P.R.China}
\affiliation{University of Chinese Academy of Sciences, 
Beijing 100049,
P.R.China}
\affiliation{Yunnan Key Laboratory of Solar Physics and Space Science, Kunming 650216, P.R.China}

\author{Tanmoy Samanta}
\affiliation{Indian Institute of Astrophysics, Koramangala, Bangalore 560034, India} 

\author{Guanchong Cheng}
\affiliation{Yunnan Observatories, 
Chinese Academy of Sciences, 
Kunming, 
Yunnan 650216, P.R.China}
\affiliation{University of Chinese Academy of Sciences, 
Beijing 100049,
P.R.China}
\affiliation{Yunnan Key Laboratory of Solar Physics and Space Science, Kunming 650216, P.R.China}

\author{Yifu Wang}
\affiliation{Yunnan Observatories, 
Chinese Academy of Sciences, 
Kunming, 
Yunnan 650216, P.R.China}
\affiliation{University of Chinese Academy of Sciences, 
Beijing 100049,
P.R.China}
\affiliation{Yunnan Key Laboratory of Solar Physics and Space Science, Kunming 650216, P.R.China}

\author[0000-0003-3439-4127]{Robert Erd\'{e}lyi}
\affiliation{Solar Physics and Space Plasma Research Centre, 
School of Mathematical and Physical Sciences, 
University of Sheffield, \\
Hicks Building, Hounsfield Road, 
Sheffield, 
S3 7RH, UK}
\affiliation{Department of Astronomy, 
E\"{o}tv\"{o}s Lor\'{a}nd University, 
P\'{a}zm\'{a}ny P\'{e}ter s\'{e}t\'{a}ny 1/A, 
Budapest,
H-1112, Hungary}
\affiliation{Gyula Bay Zolt\'{a}n Solar Observatory (GSO), 
Hungarian Solar Physics Foundation (HSPF), 
Pet\"{o}fi t\'{e}r 3,
Gyula, 
H-5700, Hungary}

\begin{abstract}

How the solar atmosphere is heated from a temperature of about $5,000-6,000$\,K in the lower atmosphere to about $1-2$\,MK in the corona has challenged the astrophysical community for about 80 years. The same puzzle exists for the stellar coronae heating as well. In this study, we present a series of findings on solar spicules and their subsequent impact on the corona within a coronal hole environment, characterized by locally open magnetic field lines, combining insights from MHD simulations with observations. We find that the convective and turbulent motions around the solar surface cause plenty of shocks and small-scale magnetic reconnection in the lower atmosphere. The combined effects of shock compression and reconnection outflows then drive the formation of groups of spicules with a quasi-period of about $300$\,s and width of $\sim 200-500$\,km. The spicule upflows provide an averaged mass flux above $10^{-9}$\,kg\,m$^{-2}$\,s$^{-1}$ in the lower corona to sustain the solar wind in coronal holes, and they continuously trigger further new local slow-mode waves and shocks. These waves supply an energy flux of $10-100$\,W\,m$^{-2}$ in the lower corona, and they are dissipated by heat conduction and compression heating to sustain the corona temperature of about $1$\,MK. The results also indicate that the upward propagating disturbances (PDs) observed in extreme ultraviolet (EUV) passbands are caused by both spicule upflows and slow-mode waves and shocks. Our findings help to understand the long standing problem of coronal heating and the origin of solar winds in coronal hole regions.

\end{abstract}

\section{Introduction} \label{sec:intro}

In the Sun's core, nuclear reactions generate immense energy. While the core reaches an extreme temperature of about $15$\,MK, it cools significantly to around $5,000-6,000$\,K at the solar surface. Inexplicably, above the lower solar atmosphere, the temperature sharply rises to about $1-2$\,MK in the corona. This puzzling behaviour, often refers as coronal heating problem, also exists for the stellar coronae. Solving this puzzle has to answer the questions of how the energy is transported up into the corona and then how it is converted into heat over the altitude range of several solar radii. Particularly, in a coronal hole, regions on the Sun where the plasmas density and temperature are lower than the surrounding, and the magnetic fields there are usually open in the corona \citep{2009LRSP....6....3C, 2005Sci...308..519T}. Ubiquitous magnetohydrodynamic (MHD) waves in the solar atmosphere are a major group of candidates for solar plasma heating \citep{2007Sci...318.1572E, 2012NatCo...3.1315M, 2019NatCo..10.3504L}. The convective and turbulent motions around the solar surface generate a copious supply of waves. The slow magnetoacoustic waves usually rapidly develop into shocks and damp in the lower atmosphere before they reach the corona \citep{1996SSRv...75..453N, 2004Natur.430..536D}, while fast magnetoacoustic modes reflect and refract as they propagate upward \citep{2002ApJ...564..508R}. Furthermore, mode coupling occurs where the sound speed equals the Alfv\'{e}n speed \citep{2003ApJ...599..626B}. Alfv\'{e}n waves caused by transverse perturbations could possibly carry enough energy up into the corona in the quiet Sun and coronal hole region \citep{2011Natur.475..477M,  2019NatCo..10.3504L}, but further studies are required to verify whether they can supply sufficient energy for heating the active region \citep{2023NatAs...7..856Y, 2021SoPh..296...47T}. How the transverse waves are dissipated in the corona is another open question \citep{2014masu.book.....P}. 

Coronal holes are considered a prominent source of the solar solar wind - a continuous stream of charged particles that escape from the lower solar atmosphere \citep{2009LRSP....6....3C}. The fast solar wind originating from coronal holes \citep{1973SoPh...29..505K, 2005Sci...308..519T} largely determines the electromagnetic field environment of the heliosphere, impacting the space environment of the Earth and the operation of satellites \citep{2018LRSP...15....1R}. The coronal hole regions may often cover a significant portion of the Sun's surface, sometimes reaching up to 30\% or more, during a solar activity minimum period \citep{2009LRSP....6....3C, 2017SoPh...292...18L}. To sustain a temperature of several hundred thousand K to $1$\,MK in the coronal hole region, an energy flux of approximately $1000$\,W\,m$^{-2}$  in the chromosphere and $70$ \,W\,m$^{-2}$ in the corona is required to balance energy losses due to heat conduction and radiative cooling \citep{1977ARAA..15..363W}.

Small-scale transient spicules are jet-like features that ubiquitously appear in the solar chromosphere and transition region \citep{Secchi1877, 2011Sci...331...55D, Pereira2012, 2014Sci...346A.315T}. The characteristics of solar spicules and jet like features observed in different wavelengths have been summarized in the comprehensive review papers \citep[e.g.,][]{2000SoPh..196...79S, 2012SSRv..169..181T, 2021RSPSA.47700217S}, the numerical models and triggering mechanisms of solar spicules have also been discussed. These elongated features connect the cool solar surface to the hot corona, and various MHD waves and Joule heating process are induced associating with the generation process of spicules. Therefore, they may play an important role in supplying energy and material to the corona \citep{1999Sci...283..810H, 2011Sci...331...55D, 2019Sci...366..890S, 2023Sci...381..867C}.

The \textit{Interface Region Imaging Spectrograph} (\textit{IRIS}) solar observation satellite \citep{2014DePontieu} and the Atmospheric Imaging Assembly (AIA) instruments \citep{2012Lemen} on board the \textit{Solar Dynamics Observatory}  \citep{Pesnell2012} have provided plenty of high quality observational data to study solar activities and the resulted perturbations. The perturbations manifest as slanted ridges of alternating brightness (or Doppler velocity) in time-distance maps, which are frequently used to analyze the development of a particular coronal structure. These perturbations are generally referred to as propagating disturbances (PDs). The combined observations of \textit{IRIS} 1400 {\AA}, \textit{SDO}/AIA 304 {\AA} and 171 {\AA} indicate that the beginning of quasi-periodical upward PDs in the coronal holes usually coincides with the rising phase of spicules in the lower atmosphere \citep{2015ApJ...809L..17J,2015ApJ...815L..16S, 2021SSRv..217...76B}. Both the slow magneto-acoustic waves and upflows connected with spicules may contribute to the generations of PDs \citep{1998ApJ...501L.217D, 2012ApJ...759..144T}. However, it is still not clear how they play their roles in generating PDs, respectively. If slow magneto-acoustic waves indeed frequently appear in the corona and cause the PDs observed in extreme ultraviolet (EUV) passbands, the problem then arising is where they are triggered, since the slow-mode waves generated around the solar surface are usually rapidly damped in the lower atmosphere. The next important question is whether these slow-mode waves can play significant roles in heating the coronal hole regions.

The early one dimensional simulations have investigated how the initiated gas pressure gradient and shock compression triggered spicules \citep[e.g.,][]{1982SoPh...78..333S, 1993ApJ...407..778S, 2004Natur.430..536D}. Then, the advanced two dimensional radiative MHD simulations further prove that the convective motions can lead to the formation of slow-mode shocks and trigger solar spicules \citep[e.g.,][] {Hansteen2006ApJL, Heggland2011ApJ,2024ApJ...973...49K}. These previous works focus on slow-mode shocks and the formation process of solar spicule below the transition region. So far, the radiative MHD studies of the corona response arising from slow-mode shocks and spicules in the corona holes are rare. In this Letter, we investigate the energy and mass transport associated with spicule upflows in coronal holes through radiative MHD simulations and observations. The triggering process of slow mode waves and shocks and their roles for corona heating in the lower corona have been analyzed in detail. We also discussed the generation mechanisms of PDs associated with spicules and compare the numerical results with observations. In the next section, we introduce the observational results. The numerical models are described in Section 3. The main results of the model analysis are presented in Section 4. We will give conclusions and discussions in Section 5.  

\section{Observational results} \label{sec:Data}
We used data obtained from coordinated observations with the IRIS satellite and the AIA instrument onboard the SDO. The IRIS observation captured a northern coronal hole region, where several polar plumes are clearly visible in AIA images (Figure 1). In this study, we used data obtained during 23:05 UT on May 17, 2020, to 01:15 UT on May 18, 2020. During this period, IRIS observed in sit-and-stare mode. We used Level 2 processed IRIS data, which has been corrected for dark current, flat-fielding, and geometric distortions. All IRIS Slit-Jaw Images (SJIs) were taken with an 8-second exposure time and a cadence of 36.88 seconds. The pixel sizes of IRIS SJIs are $0.166^{\prime\prime}$, however we rebin the image $2\times2$ pixels in Y-axis. AIA images were acquired with a 12-second cadence and have a pixel size of $0.6^{\prime\prime}$. AIA pixels are rescaled to match IRIS data for easier comparison. The IRIS SJIs and AIA images were co-aligned using IRIS 1400 {\AA} and AIA 1600  {\AA}  images, following the method described in \cite{2015ApJ...815L..16S}. For this study, we primarily used IRIS from the 2796 {\AA} SJI, which is dominated by the Mg II k  2796 {\AA}. Additionally, we analyzed AIA filtergram images centered at 171 {\AA} and 193  {\AA}, dominated by Fe IX and Fe XII emissions, respectively. The IRIS 2796  {\AA}  passband is sensitive to plasma at temperatures of approximately $10,000-15,000$\,K. The AIA 171  {\AA} and 193 {\AA} filters have peak response functions at approximately $0.8$\,MK and $1.25$\,MK, respectively.

\begin{figure}
\centerline{\includegraphics[scale=0.5,clip=true,trim=0 0 0 0]{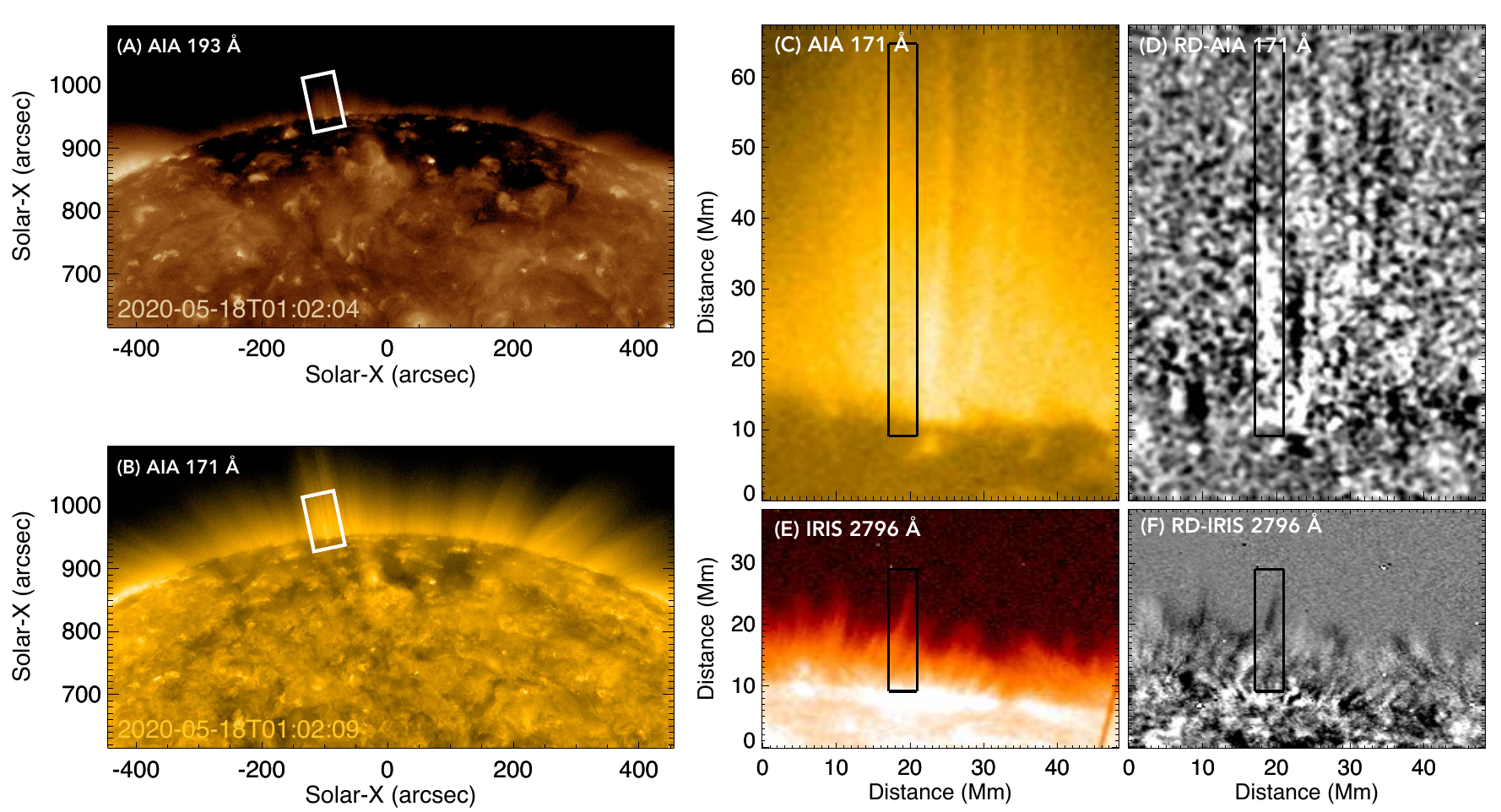}}
\caption{\textbf{AIA and IRIS images taken around 01:02 UT on May 18, 2020. } (A-B) show coronal holes in the Sun's south polar region, with the rectangular box marking the location of polar plumes. (C-D) present a zoomed-in view and a running difference image of AIA 171 {\AA}, respectively. The running difference (RD) image highlights a propagating disturbance, visible as an elongated white strip within the black box. The RD images are obtained by subtracting an image taken two minutes earlier. (E-F) display the IRIS 2796 {\AA} image and its corresponding RD image, clearly capturing a large spicule at this moment. The \textit{Y}-axis of panels (C-F) starts from the bottom of the white box marked in panels (A-B). The black rectangular boxes in panels C and E mark the location of the artificial slits used to generate time-distance plots.}
\end{figure}

Time-distance plots are widely used to analyze propagating features and their dynamics. In this study, we identified several propagating disturbances (PDs) in the AIA 171 {\AA} and 193 {\AA} running difference (RD) images (Figure 1). The RD images were obtained by subtracting an image taken two minutes earlier. To construct the time-distance maps for AIA and IRIS images, we used a wide slit, as marked in panels (C) and (E) of Figure 1. A wide slit was necessary because the signal in AIA channels is relatively weak for off-disk features, and spicules often exhibit transverse motion. To ensure the same spicule structures were captured and to improve the signal-to-noise ratio, we used slit with a width of 1.92 Mm. The time-distance maps for AIA channels were generated by averaging the intensity along the width of the slit. For AIA 171 {\AA} and 193 {\AA}, we processed the time-distance maps by removing a smoothed background trend along the time axis (30 minutes) to enhance the visibility of alternating bright and dark ridges. In contrast, for the IRIS 2796  {\AA} channel, we used the maximum intensity along each row of pixels within the slit, as the background intensity is much lower when spicules are absent.

The resulting time-distance maps are shown in Figure 2. The IRIS 2796 {\AA} time-distance maps, which are sensitive to chromospheric plasma, reveal the evolution of several spicules. Meanwhile, the AIA 171 {\AA} and 193 {\AA} maps show dark and bright ridges extending over longer distances, corresponding to PDs. The temporal evolution of spicules in the IRIS 2796 {\AA} channel exhibits substructures that rise and fall across all passbands, often following a parabolic trajectory, a characteristic feature of many spicules. To highlight the evolution of prominent spicules, we have marked their trajectories with white parabolic curves. These same curves are over plotted on the AIA 171 {\AA} and 193 {\AA} maps, illustrating the connection between spicular activity observed in the IRIS channel (bottom) and PDs seen in the AIA channels (top). Notably, the onset of PDs nearly coincides with the rise of the spicular envelope, and the subsequent descent of the spicular envelope is followed by brightenings and the generation of another upward PD in the AIA 171 {\AA} and 193 {\AA} channels.

\begin{figure}
\centerline{\includegraphics[scale=0.62,clip=true,trim=0 0 0 0]{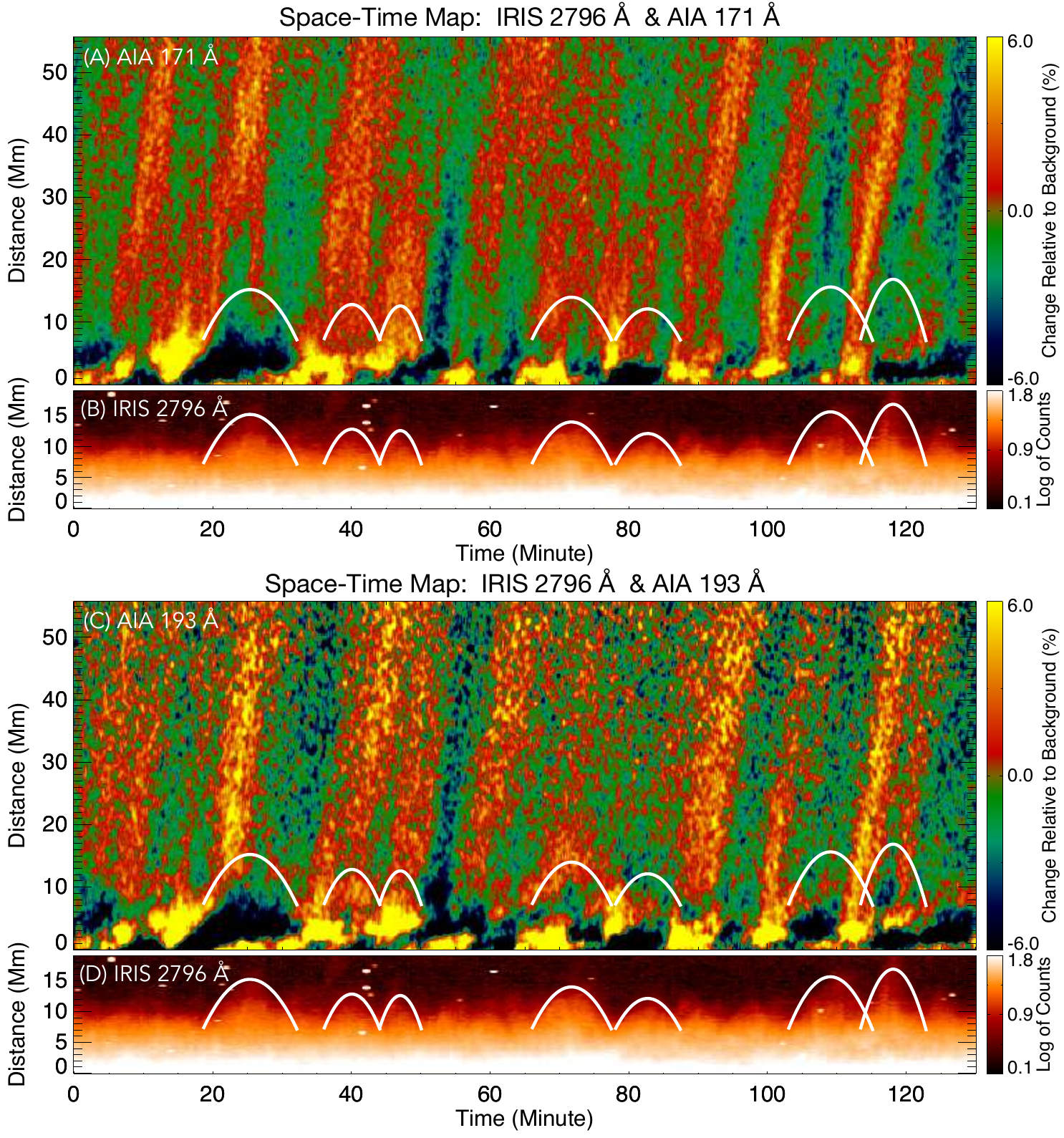}}
\caption{\textbf{Time-distance maps corresponding to the slits marked in panels (C-E) of Figure 1.} The slanted bright ridges extending over large distances represent propagating disturbances (PDs) observed in the AIA 171 {\AA} and 193 {\AA} channels. The IRIS 2796 {\AA} time-distance maps show the evolution of spicules. To highlight these evolution of several prominent spicules, we have marked spicular evolutions with white parabolic curves. These same curves are overplotted on the AIA 171 {\AA} and 193 {\AA} maps to illustrate the connection between spicular activity observed in the IRIS channel in (B) and (D) and PDs seen in the AIA channels in (A) and (C).}
\end{figure}

\section{MHD models}

In the numerical model, we assume the solar plasma is composed of hydrogen and helium. The partially ionized lower solar atmosphere includes hydrogen atoms, helium atoms, electrons, hydrogen ions and helium ions. In the fully ionized corona, the electrons, hydrogen and helium ions contribute the plasma compositions. All the plasma species are assumed to be coupled as one fluid when we solve the MHD equations, and the partial ionization effects can be included in the magnetic induction and energy equations. In this work, we have considered the magnetic diffusion contributed by both electron-ion and electron-neutral collisions, but the ambipolar diffusion is turned off. Based on the open source NIRVANA code \citep{2008Ziegler}, we have developed the code by including the partial ionization effects of both hydrogen and helium, and the radiative cooling models have also been updated. Four different radiative cooling models for the convection zone \citep{1994sse..book.....K}, photosphere \citep{2012Abbett}, chromosphere and corona \citep{2012Carlsson} are included, respectively. The initial uniform magnetic field is in the \textit{Y}-direction and vertical to the solar surface. In our recent paper \citep{2025Wang}, the details of the solved MHD equations, the formula of the physical magnetic diffusivities and four radiative cooing models applied from the upper convection zone to the corona have been presented.  

We have performed 2.5 dimensional (2.5D) MHD simulations, the numerical domain extends from $-6$\,Mm to $6$\,Mm in the \textit{X}-direction and from $-3$\,Mm to $22$\,Mm in the \textit{Y}-direction. The \textit{X}-direction is parallel to the solar surface and the \textit{Y}-direction is vertical to the solar surface. The region from $-3$\,Mm to $0$\,Mm in the \textit{Y}-direction represents the upper convection zone. In the two cases presented in Figures 4-9, the numerical resolution is $896\times2048$. We also performed a higher resolution case with grids $1792\times2048$. The grid size is $6.7$\,km in the \textit{X}-direction and $12.2$\,km in the \textit{Y}-direction, and the results for this highest resolution run are presented in Figures 10-12.     
	
The C7 model \citep{Avrett2008} and standard Solar Model \citep{1974SoPh...34..277S} are applied to set the initial plasma temperature, density and pressure above the solar surface and in the convection zone, respectively. These initial parameters along the \textit{Y}-direction are plotted in Figure 3(A). The initial magnetic field is uniform, and we have run cases with different strength of magnetic fields, e.g. $B_{y0}=B_0=5$\,G and $B_{y0}=B_0=20$\,G. All the initial variables are uniform in the \textit{X}-direction, except that we have initiated a small perturbation of the order of 0.01 in density at around $Y=-2.6$\,Mm in the convection zone to trigger the faster evolution of the whole system, which is also widely applied in most existing RMHD simulations. Such a small perturbation is applied only once at the beginning, and the quasi-periodic variation in the results is an emergent feature of the system in a self consistent manner caused by convective motions.

\begin{figure}
 \centerline{\includegraphics[scale=0.25,clip=true,trim=0 0 0 0]{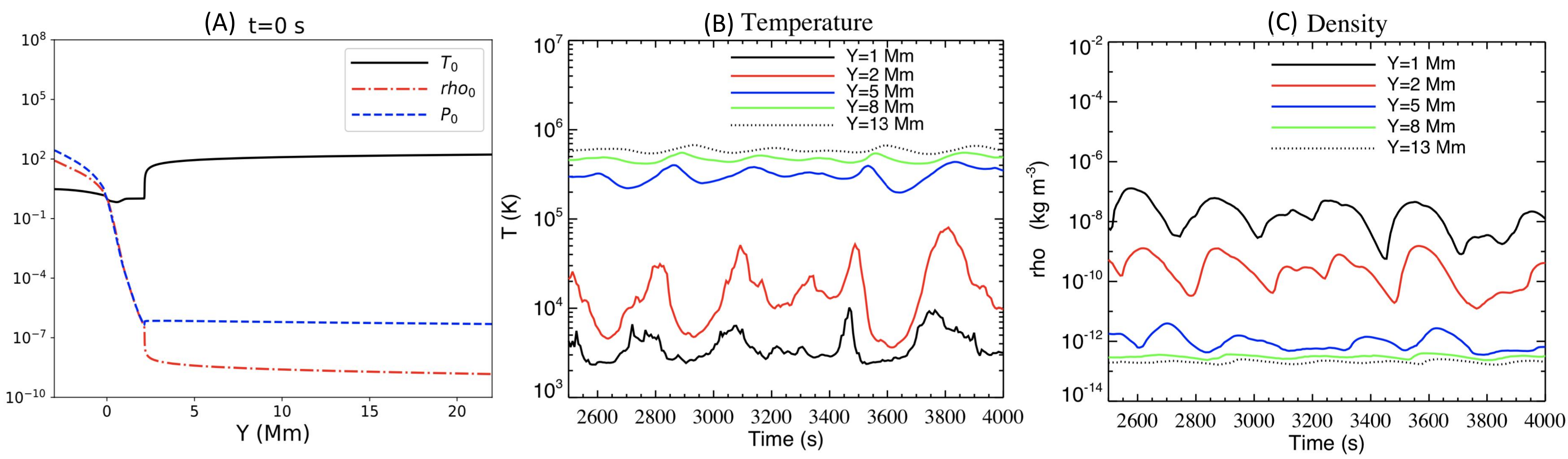}}             
\caption{ \textbf{The distributions of plasma parameters at different altitudes.} (A) shows the distributions of initial temperature ($T_0=6530$\,K), density ($rho_0=2.734 \time10^{-4}$\,kg\,m$^{-3}$) and pressure ($P_0=11820$\,N\,m$^{-2}$) along the \textit{Y}-direction, and they are normalized by using the reference values at the bottom of photosphere. The evolutions of temperature and density at different altitudes are presented in (B) and (D), and these values are the averaged ones in the \textit{X}-direction.}
\end{figure}

The horizontal boundary conditions are periodic. The bottom boundary conditions are inflow, and the up boundary conditions are outflow. At both the bottom and top boundaries, the gradients of the thermal energy, mass density, and parallel components of magnetic field are set to zero, the perpendicular component of the magnetic field is obtained by divergence-free extrapolation. The gradients of velocities in the \textit{X} and \textit{Z}-direction also vanish at the top and bottom boundaries by assuming $\partial V_x/\partial y=0$, $\partial V_z/\partial y=0$. At the bottom boundary, the fluid is only allowed to flow into the computation domain by assuming:
\begin{displaymath}
  V_{ybg} = \left\{ \begin{array}{ll}
  V_{ybl} & \textrm{if $V_{ybl}>0$}\\
  -V_{ybl}& \textrm{if $V_{ybl}<0$},\\
 \end{array} \right.
\end{displaymath}
where $V_{ybg}$ and $V_{ybl}$ separately represent the velocities at the two bottom ghost layers and at the first two layers inside the simulation domain in the \textit{Y}-direction.  We have also tried some other open boundary conditions, the plasma in the convection zone will always flow dramatically out from the bottom boundary as long as it is allowed to. Therefore, the inflow boundary at the bottom is our best choice.  At the up boundary, the fluid is only allowed to flow out the computation domain by assuming:
\begin{displaymath}
  V_{yug} = \left\{ \begin{array}{ll}
  V_{yul} & \textrm{if $V_{yul}>0$}\\
  -V_{yul}& \textrm{if $V_{yul}<0$},\\
 \end{array} \right.
\end{displaymath}
where $V_{yug}$ and $V_{yul}$ separately represent the velocities at the two up ghost layers and the last two layers inside the simulation domain in the \textit{Y}-direction. We have also added a strong magnetic diffusion coefficient $\eta_{d}=10^9[\tanh(y-20L_0)/(0.2L_0)+1]$ around the up boundary, with $L_0=10^6$\,m and $\eta_d$ is in units of  m$^2$\,s$^{-1}$. Such outflow boundary and strong magnetic diffusion near the top can prevent plasmas and waves reflect back from the up boundary. We have measured the averaged mass flux and energy flux at both the bottom and up boundaries during the period between $3000$\,s and $4500$\,s. Though the inflow boundary is applied at the bottom, there are no mass and energy flowing into the simulation domain from the bottom boundary during this period. The mass and energy continually flow out the simulation domain from the up boundary, and the average mass and energy fluxes are about $10^{-9}$\,kg\,m$^{-2}$\,s$^{-1}$ and $100$\,W\,m$^{-2}$ respectively, which are close to the values measured from the recent numerical simulation performed by using the advanced MuRaM code \citep{2025AA...702L...4C}. Such results indicate that the simulations are not influenced by external  factors and the numerical results are reliable during the period shown in this paper. However, we should point out that the convective motions obviously become weaker after $5000$\,s. In the future work, the bottom boundary with suitable inflowing mass and energy for balancing the mass and energy loss from the up boundary is required to sustain the convective motions, and we can expect that the longer spicules with faster upflows will be generated. 	    

\section{Main Results} \label{sec:Results}

\subsection{The simulated Spicule-like features}

The simulated system in all cases is initially not in equilibrium and undergoes significant evolution over time. The convective motions around the solar surface are self-consistently generated. After about $2500$\,s, the system reaches the state of dynamic equilibrium. Groups of spicules are quasi-periodically ejected upward and fall down, with a period about 300 s. Distributions of temperature, plasma density and velocity in the \textit{Y}-direction at five different times during one period are presented in Figure 4. As shown in this figure, the big thick black arrow in each panel points to the top of one such spicule with cold elongated plasma jet. It starts to rise at about $3481$\,s, reaches its maximum height at $3616$\,s, and then begins to fall (compare panels A-E and K-O of Figure 4). 

Figure 5(A) shows such a group of spicules just appear and reach a height at around $1$\,Mm above the solar surface, the maximum velocity of the cold chromospheric spicule is above $60$\,km\,s$^{-1}$ (see the area indicated by the big thick arrow in Figure 4(K)), which is close to the velocity of the faster type of spicules from observations \citep[e.g.,][] {Pereira2012, 2012ApJ...752L..12D}. During the period between $3460$\,s and $3760$\,s we have detected 50 spicules, assuming the width of the spicule is about $200$\,km, then we can get the occurring frequency of the spicule is $\sim 0.1$ spicules Mm$^{-2}$\,s$^{-1}$, which is about five times smaller than the recent measurements from observations \citep{2024ApJ...963...79L}. 

In our simulations, the ratio between up- and down-flowing spicules during this period is about 1, {which means that the rising spicules always fall back later. However, not all the up-flowing plasmas in the spicules return back to the solar surface. There is always a small fraction of the ejected material that ultimately reaches the corona. The detailed analysis and descriptions of this findings are presented in the next subsection. Based on the observational data from H$\alpha$ and Ca II K passbands, a previous work \citep{2021AA...647A.147B} identified approximately 20,000 rapid blue-shifted excursions (RBEs) and 15,000 rapid red-shifted excursions (RREs), corresponding to about  $34\%$ more RBEs than RREs. However, these blue/redshift statistics should not be equated with the occurrence of up-flows and down-flows. To date, no reliable statistical counts of up-versus down-flows have been established, primarily due to the difficulty of isolating downward motions in Doppler signals. There is one more point that needs to be noted. The down-flowing phase of the faster type spicules was normally not observed in H$\alpha$ and Ca II passbands \citep{Pereira2012}. One of the reasons is that they were heated to higher temperature during their rising phase, and they were observed to have down-flowing phases in the transition region passbands \citep{Pereira2014ApJL}.

\begin{figure}
 \centerline{\includegraphics[scale=0.63,clip=true,trim=0 0 0 0]{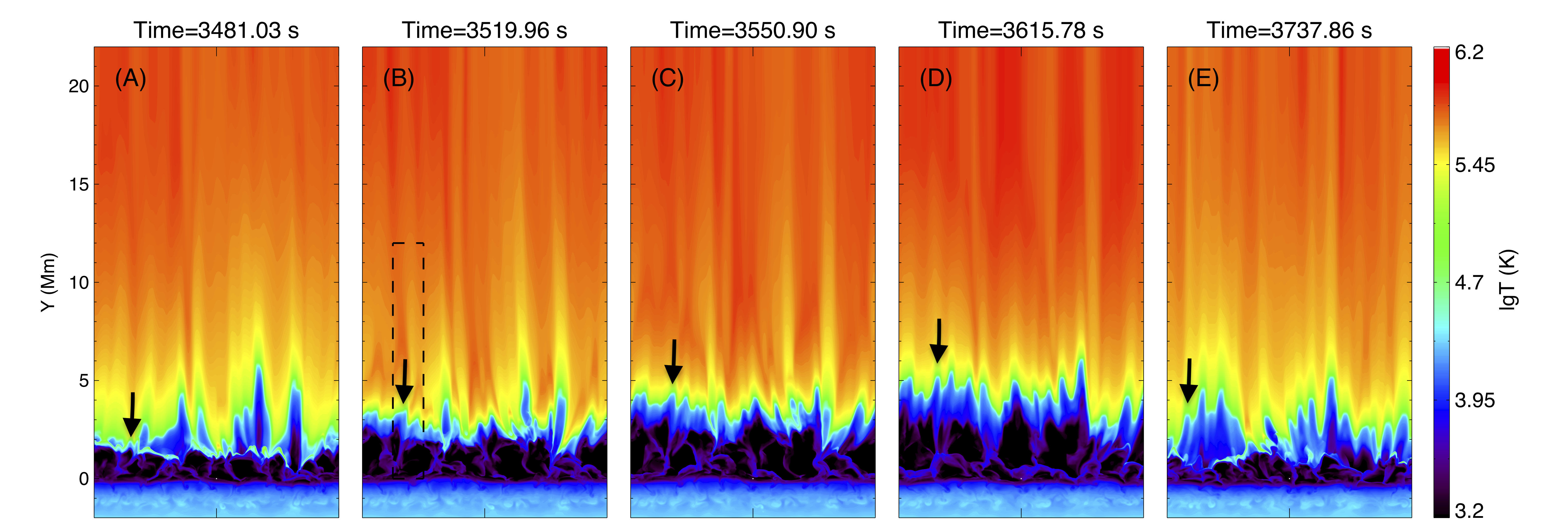}}
 \centerline{\includegraphics[scale=0.63,clip=true,trim=0 0 0 0]{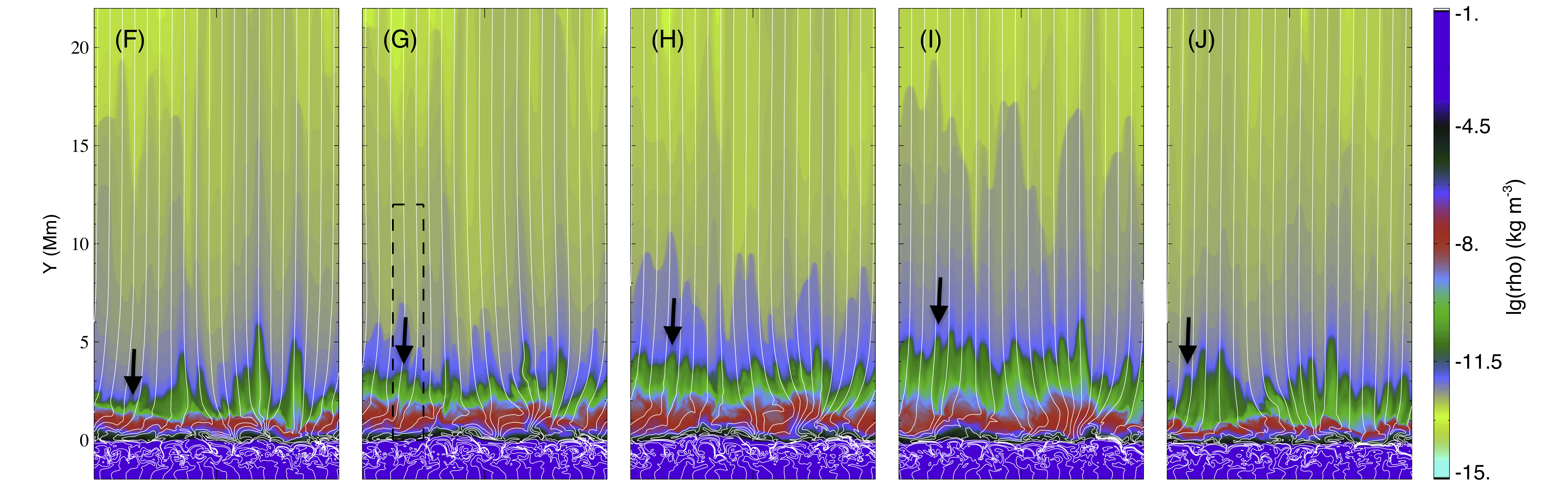}} 
 \centerline{\includegraphics[scale=0.63,clip=true,trim=0 0 0 0]{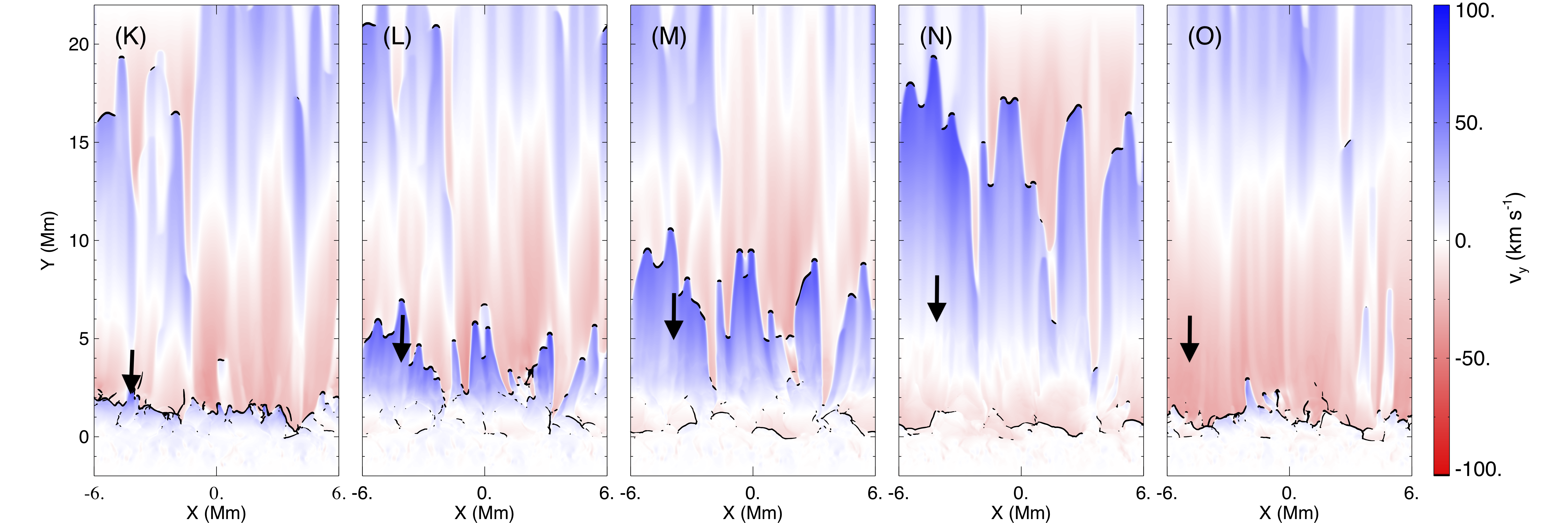}}                
\caption{\textbf{Evolution of spicules and the lower corona above them within a cycle of spicule formation in the 2.5 D simulation with the initial magnetic field $B_0=5$\,G.}  Distribution of logarithmic temperature (A-E), logarithmic density (F-J), and velocity in the $Y$-direction (K-O) at five different times are presented. The white solid curves in (F-J) outline the magnetic field lines. The black dashed boxes in (B) and (G) represent the zoomed in region in Figure 6(A-H). The big thick black arrow in each panel points to the top of one particular spicule with a temperature below 10$^5$ \,K. The black contour lines in (K-O) outline the positions having large values of $-\nabla \cdot V$, indicating the possible  invoked slow mode shocks. The animations of the corresponding temperature distribution, density distribution and velocity ($V_Y$) distribution are available in MovieS1. }
\end{figure}

The driving process of spicules highly relates to the convective and turbulent motions at around the solar surface, which result in various MHD waves \citep[e.g.,][] {1982SoPh...78..333S, 1993ApJ...407..778S, 2004Natur.430..536D}, complex curved and twisted magnetic fields \citep[e.g.,][] {2011Martinez, 2017Iijima, 2022NatPh} and magnetic reconnection processes \citep[e.g.,][] {1995Yokoyama, 2019Sci...366..890S}.  These physical processes and structures can cause the upward Lorentz force and pressure gradient to push the plasmas upward, then spicules are formed.  As shown on Figure 5 (E) and (F), we also find that the acceleration of the spicules at the primary stage is contributed by both pressure gradient and Lorentz force, but the pressure gradient is likely more dominant in most regions below these accelerated spicules.  

Figure 5(B) and 5(C) indicate that many tiny reconnection events appear below $2$\,Mm when the positive and negative magnetic fields approach each other. These tiny reconnection events change the magnetic topology in their outflow regions and can create strong local Lorentz forces, and the cumulated outflow plasmas can also give rise to the local high pressure gradients. However, the plasma environment and magnetic structures are very complicated and they vary fast with time. It is normally very difficult to distinguish whether a spicule formation process is dominated by the magnetic reconnection, the shock compression, or other mechanisms, especially when these mechanisms appear simultaneously. As shown in Figure 5(D), the shock compression is also obvious in and below the spicule formation regions. Therefore, the spicule formation process generally relates to both magnetic reconnection and shock compression in our simulations.

\begin{figure}
 \centerline{\includegraphics[scale=0.35,clip=true,trim=0 0 0 0]{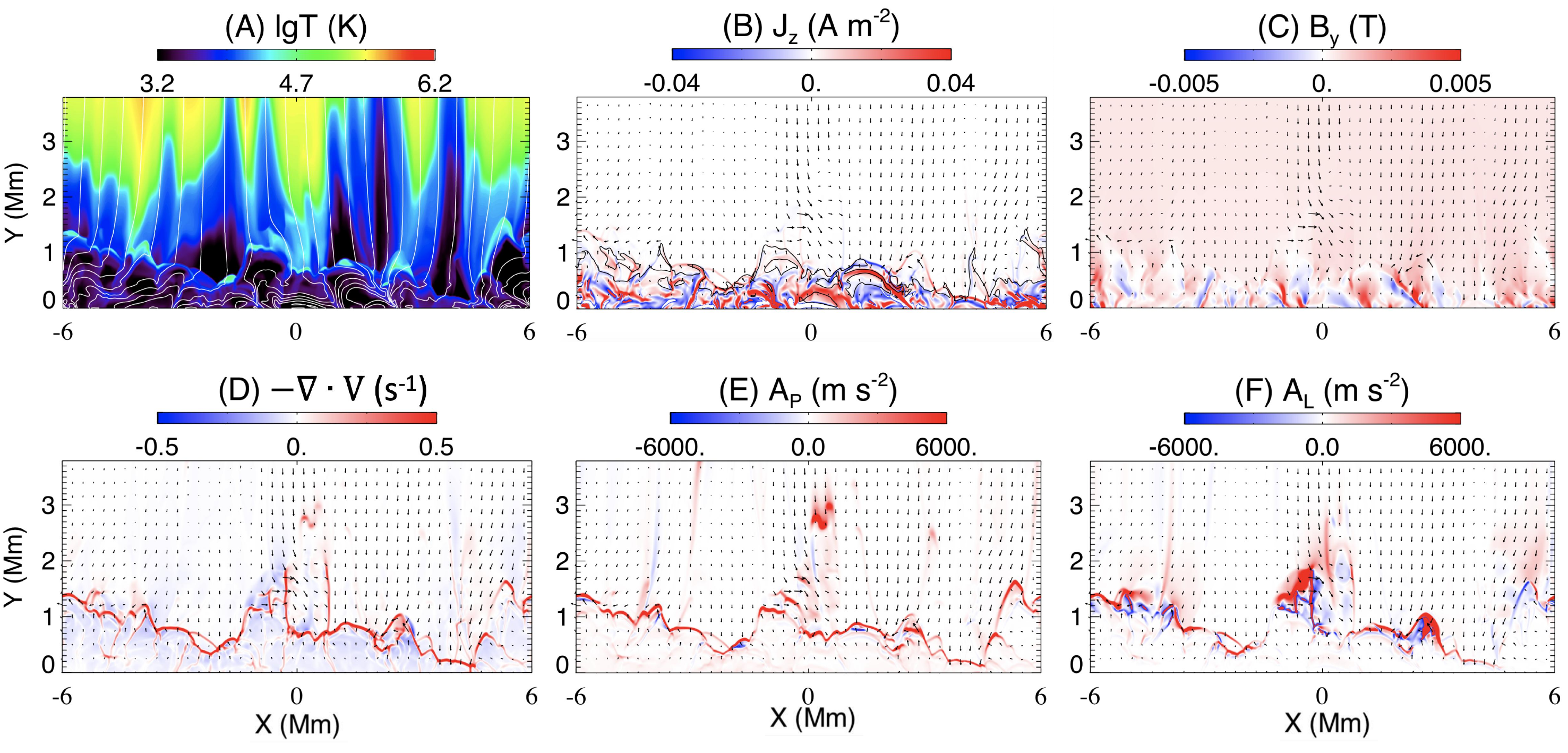}}             
\caption{\textbf{Distributions of different variables in the zoomed lower atmosphere for the 2.5 D run with $\mathbf{B_0=5}$\,G at time $\mathbf{3461.07}$\,s, when a bunch of new spicules just start to appear at the altitude between $\mathbf{Y=0.5-2}$\,Mm.} The logarithmic temperature ($lgT$) (A), current density ($J_z$) (B), magnetic field in the \textit{Y}-direction ($B_y$) (C),  the negative divergence of velocity ($-\nabla \cdot V$) (D), the acceleration along the \textit{Y}- direction contributed by the pressure gradient ($A_p$) (E),  and the acceleration along the \textit{Y}- direction contributed by the Lorentz force ($A_L$) (F) are presented. The black thin solid line in (B) represents the position where the plasma $\beta=1$.}
\end{figure}

\subsection{Mass and Energy transport associated with Spicules}

We observed that the temperature in the corona region increases during the rising phase of spicules and decreases to a lower value as the spicules fall downward (compare the temperature distributions above the spicule in the up panels of Figure 4). The averaged density and temperature at a particular height fluctuate quasi-periodically (see Figure 3(B) and 3(C)), and so does the maximum temperature. During the rising phase of these spicules, a small fraction of cool plasmas in the low atmosphere is ejected upward into the corona and heated to the coronal temperature (see MovieS1 referring Figure 4), the maximum speed of the rising plasma can reach up to $100$\,km\,s$^{-1}$ (see Figure 4(K-O)). In the later stage of a characteristic spicular formation period, a fraction of the heated plasma continues to move upward in the corona, with some flowing out through the upper boundary (see Figure 4 and the referring MovieS1), while the nether cooler components are falling back to the solar surface. 

Figure 4(K-O) show that the slow-mode shocks (the regions with large values of $-\nabla \cdot V$ outlined by black contour lines) are generated both in the lower solar atmosphere and in the corona at much higher altitudes. Figure 6(A-H) demonstrate that the slow-mode shocks are locally triggered by the spicular upflows from the lower atmosphere to the corona. Passing through the shock front, one may notice that the distribution of variables such as pressure and velocity in the \textit{Y}-direction (see Figure 6(I-J)) sharply jump across the shock fronts. The shock front above the spicule is always at the same location as that of the top of the high-speed upflows (see Figure 4(K-O)).  Since the slow-mode shock is a manifestation of the slow-mode waves in the nonlinear stage, the yet unobservable slow mode wave in the linear stage should also exist and results from perturbations of upflows in the corona. In the previous works \citep[e.g.,][]{2004Natur.430..536D, 2025ApJ...989...39S}, the slow-mode waves or shock waves were assumed to be triggered around the solar surface, then they propagate upward into the corona. Here, we show that they can also be directly triggered in the corona, offering a new perspective that addresses the limitation of slow-mode waves and shocks generated in the lower atmosphere, which are believed to be strongly damped before reaching coronal heights.

\begin{figure}
  \centerline{\includegraphics[scale=0.55,clip=true,trim=0 0 0 0]{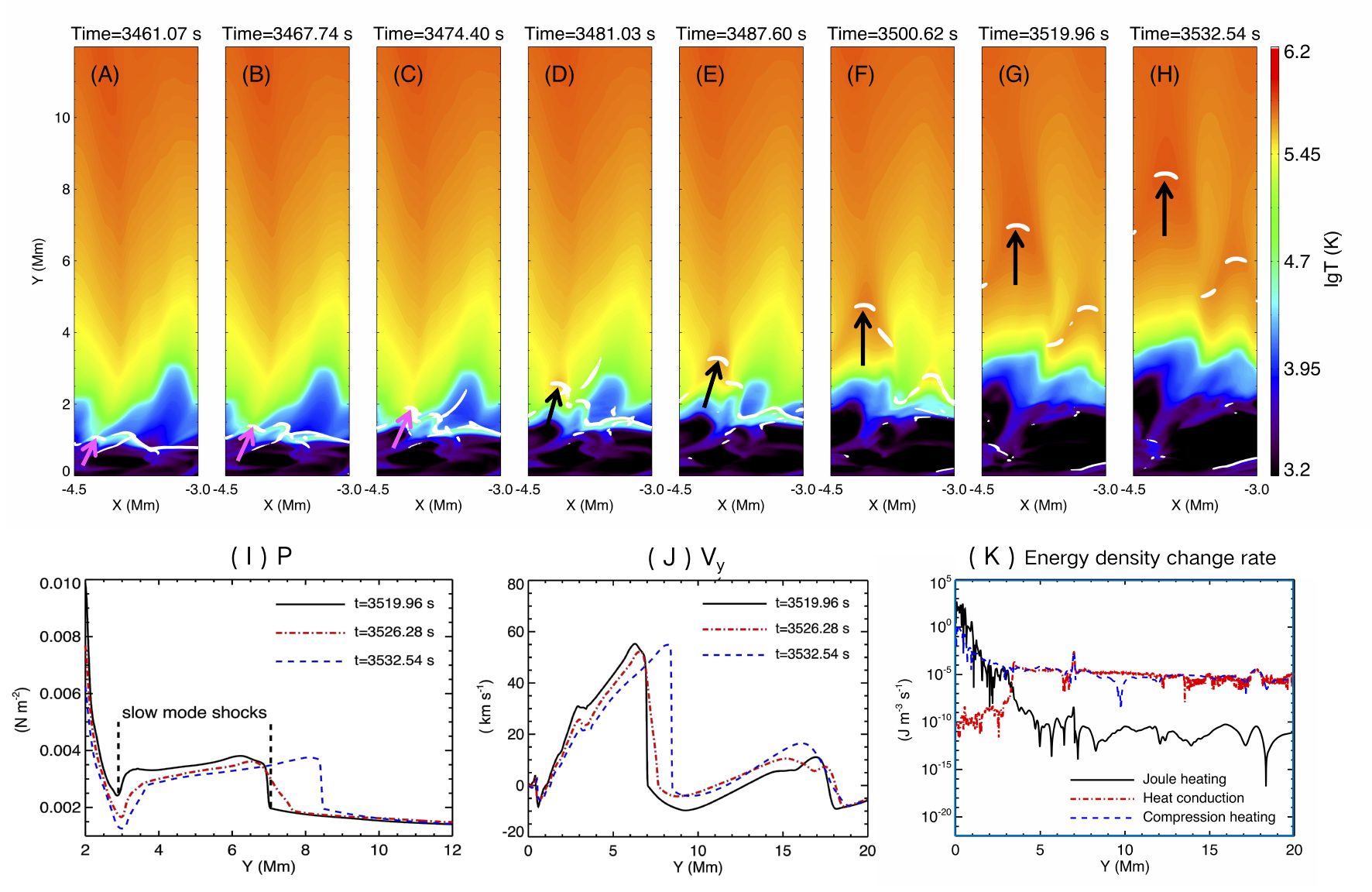}}                                    
\caption{\textbf{The slow mode shocks above the spicules in the lower corona.} (A-H) show the temperature distributions at eight different times in the zoomed in region inside the black dashed box in Figure 4(B) and 4(G), where the white contour lines outline the positions having large values of $-\nabla \cdot V$ (shock fronts), the thick pink or black arrows point to the shock front that triggered by plasma upflows during the rising phase of a growing spicule on the left hand side, and these thick arrows also represent the spicule upflows below the shock front. The distributions of gas pressure $P$, velocity in the $Y$-direction $V_y$, and the absolute values of the energy density change rate along the $Y$-direction at $X=-4.15$\,Mm are presented in (I), (J) and (K). The black solid line, the red dashed-dotted line and the blue dashed line in (I) and (J) mark the values at three different times when the shock fronts are located in the corona. The various lines in (K) indicate the absolute values of the energy density change rate contributed by the Joule heating ($\eta J_z^2$), the heat conduction ($-\nabla \cdot F_c$) and the compression heating ($-P \nabla \cdot V$) at $t=3519.96$\,s, respectively.}
\end{figure}

In addition to the slow-mode waves, other kinds of MHD waves (such as Alfv\'{e}n and fast mode waves) are also generated in the solar atmosphere. In Figure 7(A-H), the average upward energy fluxes at different altitudes, carried by these waves are measured on the basis of the differences relative to the initial quantities at the starting time of a spicule formation period \citep{2016Kanoh} (see Appendix A for details). In the chromosphere, the kinetic energy flux ($F_{kin}$) could be larger than the wave energy flux, but $F_{kin}$ decreases fast with height. In the case of a stronger background magnetic fields ($B_0=20$ G), the upward energy fluxes carried by the Alfv\'{e}n and fast-mode waves ($F_A$ and $F_{fast}$) are more than that carried by slow-mode waves ($F_{slow}$) in the chromosphere. The total average energy flux carried by three distinct MHD waves during a spicule formation period is above $1000$\,W\,m$^{-2}$ in the chromosphere in both cases (see Figure 7(D) and 7(H)). In the corona, $F_A$ and $F_{fast}$ still quickly decrease with altitude, but the energy flux carried by the the slow-mode wave ($F_{slow}$) decreases more slowly. 

Eventually, our finding is that $F_{slow}$ plays the dominant role in the context of carrying upward energy flux in the low corona. In the case of a stronger background magnetic field ($B_0=20$\,G), the maximum value of $F_{slow}$ exceeds $200$\,W\,m$^{-2}$ at $Y=13$\,Mm. During the formation period of spicules from time $3480$\,s to $3740$\,s in this case, the average value of $F_{slow}$ is around $100$\,W\,m$^{-2}$ in the lower corona, which is sufficient to balance the energy loss by heat conduction and radiative cooling in the coronal hole \citep{1977ARAA..15..363W}. In the case of a weaker background magnetic fields, $B_0=5G$, the average upward energy flux is smaller, but it is still sufficient to sustain a temperature of $\sim0.6$\,MK for the coronal hole (see panels A-E of Figure 4).  

\begin{figure}
    \centerline{\includegraphics[scale=0.50,clip=true,trim=0 0 0 0]{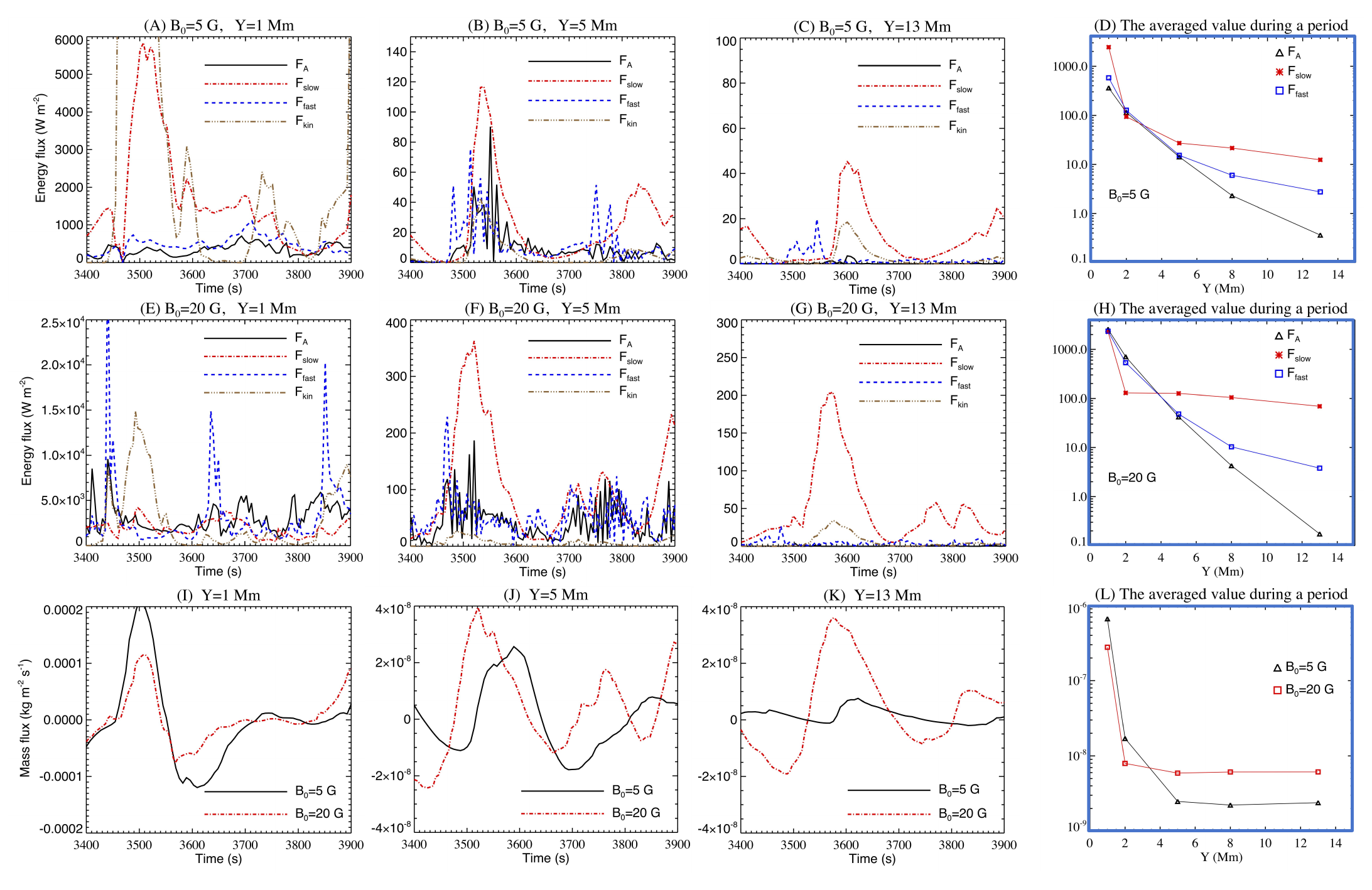}}       
\caption{\textbf{The upward energy fluxes and mass fluxes from the chromosphere to the corona.} (A-C) show the evolutions of upward energy fluxes carried by Alfv\'{e}n ($F_A$) waves, fast-mode waves ($F_{fast}$) and slow-mode waves ($F_{slow}$), and the upward kinetic energy flux ($F_{kin}$) at $Y=1$\,Mm, $Y=5$\,Mm and $Y=13$\,Mm, respectively, in the 2.5 D run with initial magnetic field $B_{0}=5$\,G. The averaged values within a cycle of spicule formation at different altitudes for this case are presented in (D). (E-G) are the same but with initial magnetic field $B_{0}=20$\,G, and the averaged values within a cycle of spicule formation at different altitudes for this case are presented in (H). The mass fluxes varying with time at $Y=1$\,Mm, $Y=5$\,Mm and $Y=13$\,Mm in the two cases are presented in (I-K), and the averaged values at different altitudes within a cycle of spicule formation for the two cases are presented in (L).}
\end{figure}

The absolute values of the three heating terms evaluated along the \textit{Y}-axis through a spicule (Figure 6(K)) and the distributions of these heating terms above the solar surface (see Figure 8) indicate that the Joule heating dominates over the other heating mechanisms in the lower solar atmosphere, but it decreases sharply with height. Eventually, the contribution from the Joule heating is overturned by those from heat conduction and compression heating by several orders of magnitude in the corona. Compression heating and  heat conduction are two major mechanisms such that they are usually considered efficient in dissipating shocks and slow mode waves \citep{2021SSRv..217...76B}. The magnetic field in the corona is almost unipolar, which causes the absence of the electric current and magnetic reconnection there (see Figures 5(B) and 8(A-E)). This result is different from previous simulations about spicule dynamics along the loops in the quiet region on the Sun \citep{2018ApJ...860..116M}, which shows that the Joule heating is significant even in the solar corona. 

\begin{figure}
 \centerline{\includegraphics[scale=0.55,clip=true,trim=0 0 0 0]{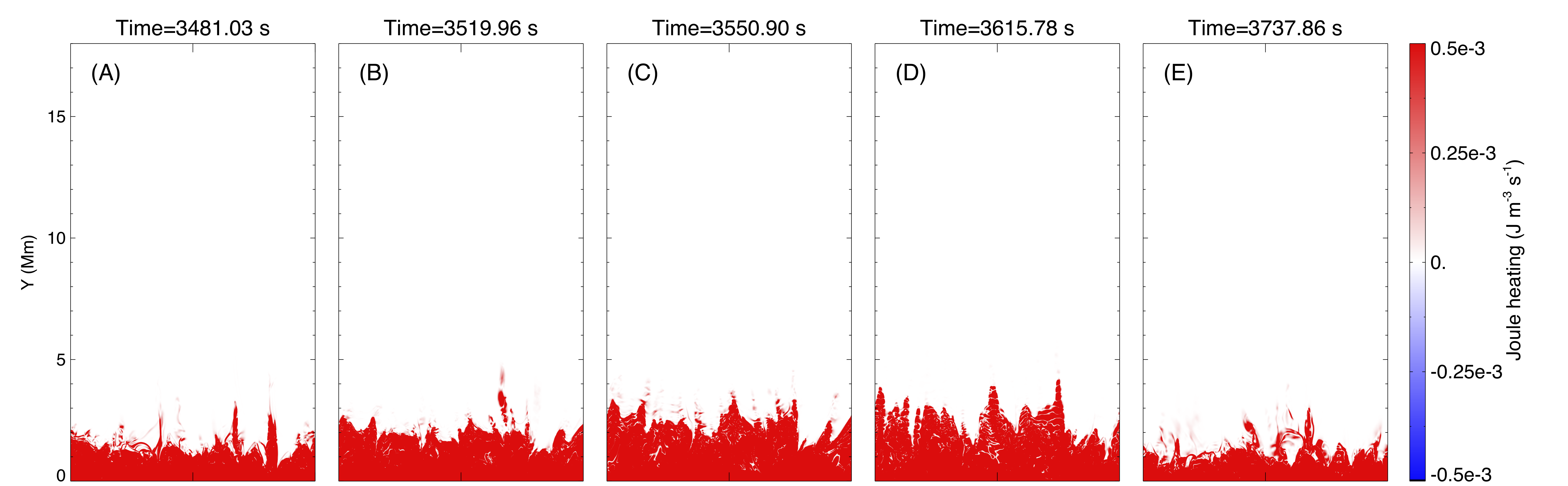}}
 \centerline{\includegraphics[scale=0.55,clip=true,trim=0 0 0 0]{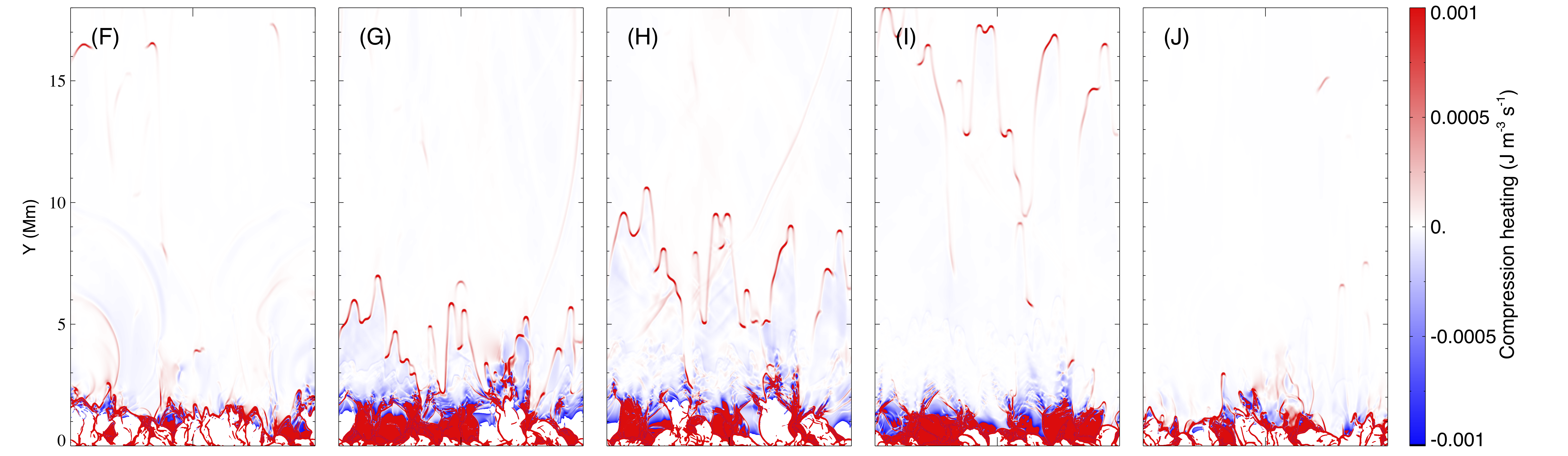}} 
 \centerline{\includegraphics[scale=0.55,clip=true,trim=0 0 0 0]{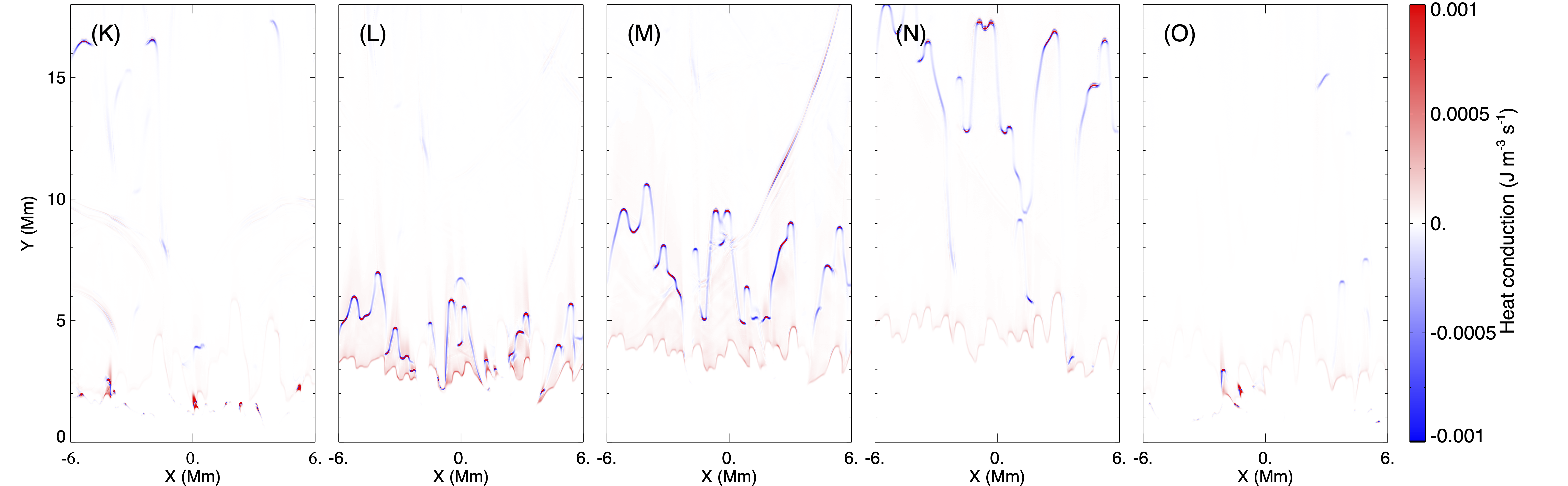}}                
\caption{\textbf{Evolutions of the energy density change rate contributed by three different physical effects within a cycle of spicule formation in the 2.5 D run with $\mathbf{B_0=5}$\,G.} Distributions of the Joule heating ($\eta J_z^2$) (A-E), compression heating ($-P\nabla \cdot \mathbf{V}$) (F-J) and heat conduction ($-\nabla \cdot \mathbf{F_C}$) (K-O) are presented. }
\end{figure}

The evolution of the average mass flux at different altitudes in the two cases are presented in Figure 7(I-L). The plots show that the plasmas are ejected upward and then most of the rising plasmas eventually fall downward within a cycle of spicule formation. However, as stated in the first paragraph of this subsection, there is always a small fraction of the ejected plasmas that ultimately reach the corona. The mass flux  dramatically drops with altitude in the chromosphere, but it only slightly decreases with altitude in the corona ($\sim$ 5 Mm above the solar surface), especially in the case of stronger magnetic field. The mass flux in the chromosphere is higher in the case of a weaker magnetic field, and the situation reverses in the corona. For example, during the period between time $3480$ s and $3740$ s, the average mass fluxes at $Y=13$ Mm are about $2.25\times 10^{-9}$ kg\,m$^{-2}$\,s$^{-1}$ and $6.12\times 10^{-9}$ kg\,m$^{-2}$\,s$^{-1}$ in the cases of $B_0=5$\,G and $B_0=20$ G, respectively. These values are consistent with the mass flux measured in the lower corona of the coronal hole regions \citep{2009LRSP....6....3C}.

\subsection{Correlations between PDs and Spicules}

It is often debated whether the coronal PDs, a commonly observed phenomena in coronal hole regions are plasma flows or slow-mode magneto-acoustic wave. Based on the numerical results, we have synthesized the emission count rates in AIA 193  {\AA} and 171  {\AA} and the results are presented in Figure 9, which shows the clear upward PDs. Comparing Figure 4(K-O) and Figure 9(A-E), one can find that the locations of the bright 193  {\AA} emission region match perfectly to the regions of upflows. Figure 9(H) shows that the enhanced emission in 193  {\AA} of PDs  always corresponds to the enhanced plasma density, and the plasma temperature also increases in most of these regions. The enhanced density is attributed to the upflows (see the region inside the black dashed box in Figure 4(G) and 4(L)), and the increased temperature is caused by the shock compression heating (see Figure 4(B) and Figure 6(G)) together with the dissipation of slow mode waves in the corona. These results clearly demonstrate that both the upflows and the slow-mode waves contribute to the formation of PDs. When the cool spicules start to slow down and then fall back, part of the EUV emissions just above these spicules also follows the similar path (see Figure 9(F) and 9(G)). However, we can still see obvious EUV emissions that move upward together with the hot upflows in the corona (see Figures 4 and 9). Therefore, though most of the ejected plasma eventually fall back, a small fraction still continuously moves upward and trigger the newly formed slow-mode waves and shocks, which then cause the upward PDs. 

Such a scenario is consistent with the observational results from IRIS and AIA. The evolution of spicules, highlighted with parabolic curves, clearly illustrates that spicules and coronal PDs are generated simultaneously, similar to what we observe in the simulation, as shown in Figure 2. The speed of upward PDs measured from observations is in the range of $100-150$\,km~s$^{-1}$. According to the results in Figure 9(F) and 9(G), the speed of upward PDs from numerical simulation is about $100$\,km~s$^{-1}$. We note that parts of the enhanced emissions in AIA 171{\AA} and 193 {\AA} might also have the rising and falling trajectory for some spicules. For example, the spicule in the middle of Figure 2 has a maximum height at about 72 minute and parts of the enhanced emissions in 171{\AA} and 193 {\AA} look like following the similar trajectory as the IRIS 2796 {\AA}, which is similar as that shown in the synthesized image in Figure 9(F) and 9(G) from numerical simulations.

\begin{figure}
 \centerline{\includegraphics[scale=0.645,clip=true,trim=0 0 0 0]{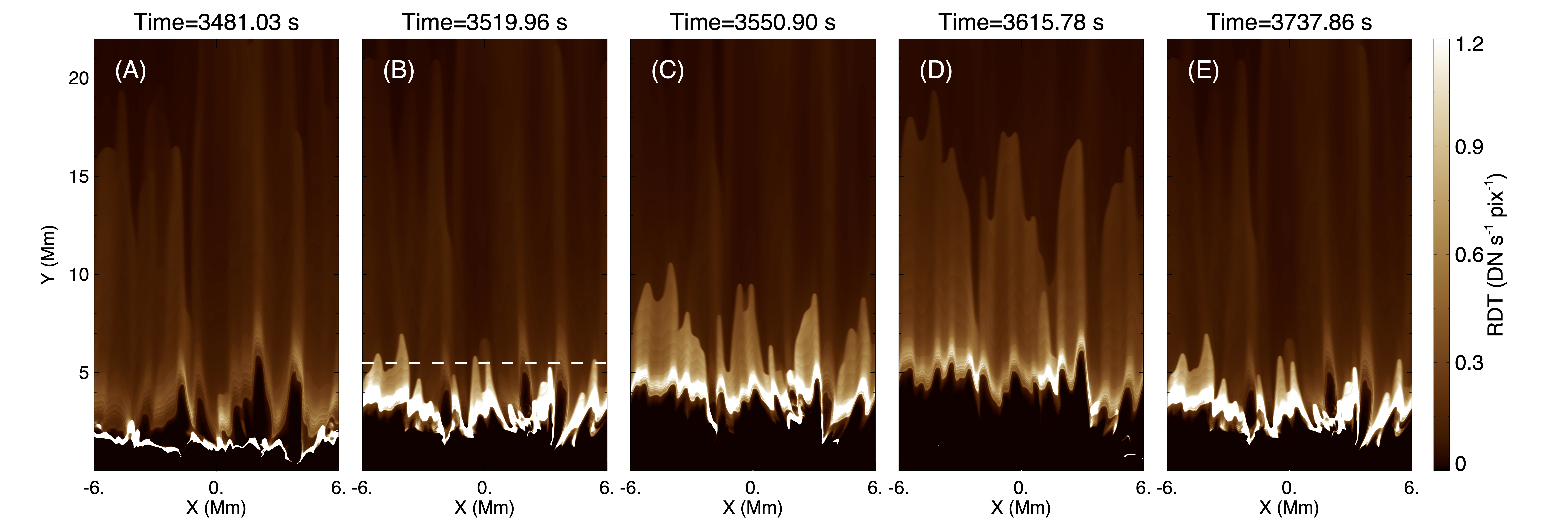}}
  \centerline{\includegraphics[scale=0.49,clip=true,trim=0 0 0 0]{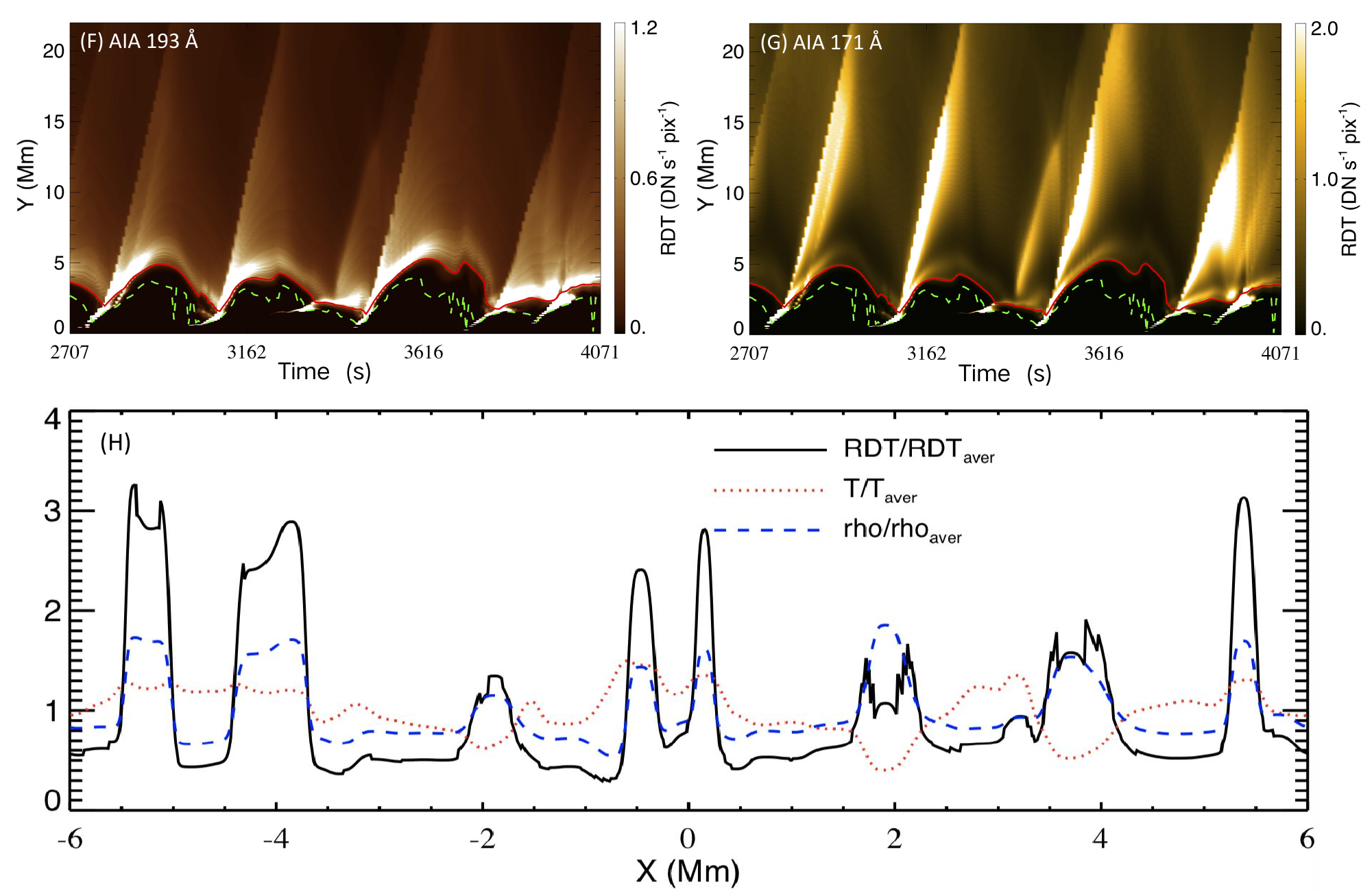}}
\caption{\textbf{Evolution of the synthesized emission count rate in the 2.5 D run with initial magnetic field $B_0=5$\,G and the observation of PDs .} (A-E) shows the distributions of the synthesized emission count rate in AIA 193 {\AA} at five different times within a cycle of spicule formation as in Figure 4. The animations of the corresponding synthesized emission count rate in AIA 171 {\AA} and AIA 193 {\AA} are available in MovieS2. (F) and (G) show the synthesized results varying with time along the \textit{Y}-direction at $X=-4.15$\,Mm. Here, $RDT$ is the synthesized emission count rate, and the unit of $RDT$ is DN\,s$^{-1}$\,pixel$^{-1}$. The red solid curve and the green dashed curve in (F) and (G) correspond to the two locations where $T=80,000$\,K and $T=8000$\,K, respectively. (H) presents the distributions of the synthesized emission count rate in AIA 193 (black solid line), the plasma temperature (red dotted line) and density (blue dashed line) along the X-direction at $Y=5.5$\,Mm (the white dashed line in Figure 9(B)) at $t=3519.96$\,s.}
\label{fig:example} 
\end{figure}

\section{Conclusions and Discussions}

MHD waves supplying energy for heating the solar corona are normally considered to be triggered by the perturbation of plasma and magnetic field in the convection zone or in the lower solar atmosphere. The slow-mode waves generated at such heights usually rapidly develop into shocks and damp in the chromosphere or the transition region. Therefore, the slow-mode waves were not considered as a plausible mechanism for heating the corona. However, the observed PDs and upflows indicate that the slow-mode waves might frequently occur in coronal holes \citep{2021SSRv..217...76B, 2021SoPh..296...47T}. This work depicts a comprehensive physical scenario about how the convections invoke the formation of spicules, upflows, slow-mode waves and shocks and how these perturbations are dissipated to heat the coronal hole regions. The main conclusions are as below:

1. The convective and turbulent motions around the solar surface produce multiple shock compression structures and  tiny small reconnection events in the lower solar atmosphere, and they both contribute the formation of solar spicules. The total acceleration is caused by both the plasma pressure gradient and the upward Lorentz force, but the plasma pressure gradient dominates. 

2. A small fraction of the upwardly ejected spicular flows can reach the corona. These spicule upflows continue to rise in the corona and some portions even flow out of the simulation domain through the upper boundary. They supply a mass flux exceeding $10^{-9}$\,kg\,m$^{-2}$\,s$^{-1}$ in the lower solar corona, that is enough to sustain the solar wind in coronal holes  \citep{2009LRSP....6....3C, 2022ApJ...940..130D}.  

3. The spicular upflows reaching the low corona continue to trigger local slow-mode waves and shocks there.  These waves are different from those originating by convection motions at around the solar surface, which can be easily dissipated before reaching corona. These waves provide an upward energy flux of $10-100$\,W\,m$^{-2}$ in the lower corona, and they can be dissipated by heat conduction and compression to heat the original cool spicule plasmas. It is worth noting that the spicule upflows appear quasi-periodically in the corona, followed by a trail of slow mode and shock waves with concurrent mass flux supplies. Such results demonstrate that slow mode waves and shocks can play an important role in heating the lower corona in the coronal hole regions. 

4. The generation of PDs in coronal holes can be attributed to both local enhanced density and temperature. The enhanced density is resulted by upflows, and the increased temperature is caused by the shock compression heating together with the dissipation of slow mode waves in the corona. 
  
According to the previous radiative MHD simulations, the PDs along loops in the quiet Sun region also associate with both shock waves and spicular flows \citep{2017ApJ...845L..18D}. In this work, we more clearly demonstrate how the PDs in the coronal hole region are caused by upflows, slow-mode waves and shocks. The slow-mode shocks or magnetic reconnection originating by convection motions in the lower atmosphere cause the formation of spicules, then the spicule upflows further compress the plasma above them to generate the new slow-mode shocks in the corona. These newly generated shocks in the corona are not those triggered by convective motions. However, in the previous papers \citep[e.g.,][]{1993ApJ...407..778S, 2004Natur.430..536D}, the slow-mode waves and shocks appearing in the corona are always assumed to be originally triggered around the solar surface or the lower atmosphere and then they propagate into the corona. Based on the radiative MHD simulations, we have investigated the mass flux and the energy fluxes contributed by slow-mode waves and shocks in the corona hole region. In addition, we find that the Joule heating is much less important than the compression heating and heat conduction in the lower corona of the coronal holes. However, the previous simulations of solar spicules along the loops in the quiet Sun region \citep{2018ApJ...860..116M} indicate that the Joule heating is still significant in the corona. The different magnetic structures and strength of magnetic fields are probably the main reasons to cause such differences.  

The wave and shock signatures and their evolutions relating with solar spicules are difficult to be distinguished. Better analytic approaches and higher resolution ground-based (e.g., Daniel K. Inouye Solar Telescope \citep{2021SoPh..296...70R}) and space solar telescopes in various wave bands are crucial for capturing the wave signals in this region. We also note the limitations of this work. The twisting motion of spicules \citep[e.g.,][]{2009SSRv..149..355Z, 2012ApJ...752L..12D, 2014Sci...346D.315D} is not considered in the present study. Therefore, the energy flux carried by the Alfv\'{e}n waves obtained here could be considered as a lower limit. The previous radiative MHD simulations proved that the ambipolar diffusion (the decoupling of ions and neutrals) can amplify the Lorentz force to cause longer and faster spicules \citep{2017Sci...356.1269M}. We expect that the ambipolar diffusion can also help to eject more plasmas with higher speed into the corona, further strengthening the formations of slow-mode and shock waves in the corona above the spicules. However, the ambipolar diffusion effect is probably weaker in the coronal hole regions with weaker magnetic field. It is necessary to check whether this effect can play an important role in the coronal hole spicule formation process in the future work.
      
There are a substantial number of sun-like stars (G-type) and stars with a mass smaller than the Sun, for example M-type and K-type stars, in the Universe. These stars have full convectional outer layers or obvious convection zones. Analogous to the Sun, these stars may produces numerous spicular type jets in their atmosphere as well. The plasma upflow from the lower atmosphere through spicules is also able to generate the slow mode wave and shock, and may play an important role in supplying mass for the stellar wind and heating the stellar coronae.

\acknowledgments

We would like to thank the referee for the comments and suggestions made for improving the quality of this work. Lei Ni thanks Prof. Hui Tian from Peking University for the discussions about spicules and PDs and Dr. Yajie Chen from Max-Planck Institute for the discussions about boundary conditions. This research is supported by the Strategic Priority Research Program of the Chinese Academy of Sciences with Grant No. XDB0560000; the National Key R\&D Program of China No. 2022YFF0503804(2022YFF0503800) and No.2022YFF0503003 (2022YFF0503000); the NSFC Grants 12373060 and 11933009; the Basic Research of Yunnan Province in China with Grant 202401AS070044; China's Space Origins Exploration Program; the Yunling Scholar Project of the Yunnan Province and the Yunnan Province Scientist Workshop of Solar Physics; Yunnan Key Laboratory of Solar Physics and Space Science under the number 202205AG070009; We benefit from the discussions of the ISSI-BJ Team "Solar eruptions: preparing for the next generation multi-waveband coronagraphs". This work was also supported by the International Space Science Institute project (ISSI-BJ ID 24-604) on "Small-scale eruptions in the Sun". R.E. acknowledges the NKFIH (OTKA, grant No. K142987) Hungary for enabling this research. R.E. is grateful to Science and Technology Facilities Council (STFC, grant No. ST/M000826/1) UK, PIFI (China, grant No. 2024PVA0043) and the NKFIH Excellence Grant TKP2021-NKTA-64 (Hungary).The simulation work was carried out at National Supercomputer Center in Tianjin, and the calculations were performed on Tianhe new generation supercomputer. The numerical data analysis have been done on the Computational Solar Physics Laboratory of Yunnan Observatories.

\appendix

\section{Methods for calculating the energy fluxes} 

In the simulation domain, the wave energy flux vector can be measured as below:
\begin{equation}
  \mathbf{F}_{wave}=\tilde{P} \mathbf{v}+ \frac{1}{\mu_0} (\mathbf{B}_b \cdot \tilde{\mathbf{B}}) \mathbf{v}-\frac{1}{\mu_0}(\mathbf{v} \cdot \tilde{\mathbf{B}}) \mathbf{B}_b,
\end{equation}
where the subscript b represents the background variable, a tilde represents the perturbation from the background conditions, and $P$ represents the plasma pressure \citep{2015Mumford}. This equation has been widely used and discussed in the previous papers \citep[e.g.,][]{2003ApJ...599..626B, 2015Mumford}. In this work, we also employ it to calculate the averaged energy fluxes along the $y$-direction, the  energy fluxes carried by the slow mode wave ($F_{slow}$), fast mode wave  ($F_{fast}$), Alfv\'{e}n wave ($F_A$), and the kinetic energy flux $F_{kin}$ along the $y$-direction at a particular height can be simply derived as below:

\begin{equation}
  F_{slow}= \frac{\int \frac{B_{0y}}{B_0} \mathbf{V}_{\parallel} \delta P dx}{\int dx},
\end{equation}

\begin{equation}
  F_{fast}=\frac{1}{\mu_0} \frac{\int B_{0y} \delta B_{\parallel} V_{\perp}dx}{\int dx},
\end{equation}

\begin{equation}
  F_{A}=\frac{1}{\mu_0} \frac{\int B_{0y}\mathbf{V}_{\perp} \cdot \delta \mathbf{B}_{\perp} dx}{\int dx},
\end{equation}

\begin{equation}
  F_{kin}= \frac{\int \rho V^2 V_y dx}{\int dx}.
\end{equation}
In the above equations, $\delta \mathbf{B} = \mathbf{B}-\mathbf{B}_0 $ and $\delta P =  P-P_0 $. The subscripts $0$, $\perp$ and $\parallel$ represent the initial values and the values perpendicular and parallel to the initial magnetic fields, respectively.  We used the values at the beginning of a spicule formation period as the initial ones to calculate the related upward (positive) energy flux as shown in Figures 7(A-H), 11 and 12.  The similar methods for determining the energy flux carried by MHD waves have been used in previous works \citep{2016Kanoh, 2020Kotani}. The mass flux shown in Figure 7(I-K) and Figure 11(E-H) is calculated as:

\begin{equation}
  F_{mass}= \frac{\int V_{y} \rho dx}{\int dx}.
\end{equation}

\section{Synthetic plasma emission}

The synthetic emission count rates in AIA 171 {\AA} and 193{\AA} as shown in Figure 9 are given by:
\begin{equation}
  RDT =  \int n_e^2 f(T) dz,
\end{equation}
 where $n_e$ is the electron density and $f(T)$ is the temperature dependent response function \citep{2012Lemen} corresponding to different wavelengths in AIA  from the Chianti package \citep{2015DelZanna}. The unit of $RDT$ is DN\,s$^{-1}$\,pixel$^{-1}$. In our 2.5D simulations, all the variables including $RDT$ are uniform in the \textit{Z}-direction and we assume the length scale in z-direction is $1$\,Mm, that is comparable to the size of a spicule.  

\section{The effects of numerical resolutions}

In order to check the effects of resolutions on numerical results and conclusions. We have performed simulations with both a lower and a higher resolutions for the case with an initial magnetic field of $5$\,G. The numerical results show that the width of the spicules is narrower and the length can reach a higher value in the simulation with a higher resolution. When the grid size is $12.2$\,km in both \textit{X} and \textit{Y} direction, the width of the spicule is in the range $\sim200-500$\,km and the maximum length can reach $\sim8$\,Mm (see Figure. 4). The width of the spicule can reach a very small value of $\sim120$\,km and the maximum length is above $10$\,Mm (Figure 10) in the simulation with an extreme small grid size of $6.7$\,km. The maximum temperature and velocity in the corona region in the higher resolution simulation are both higher, and the corona temperature is also more nonuniform in this simulation (comparing Figures 4(A-E) and 10(A-E)). 

According to the methods presented in the Appendix A and equations from A1 to A6, we have calculated the average energy fluxes carried by Alfv\'{e}n ($F_A$), fast-mode ($F_{fast}$) and slow-mode ($F_{slow}$) waves, and the mass flux at various altitudes for the simulations with different resolutions. The values of the averaged upward energy fluxes, mass flux and their trends of changing with time and altitudes in the higher resolution simulation (Figures 11 and 12) are close to those from the lower resolution simulation (compare Figures 7, 11 and 12). The upward energy flux carried by slow mode waves also dominates in the corona region in the higher resolution simulation, as shown in Figures 11(A-D) and 12(A). Therefore, our conclusions are not affected by numerical resolution. However, comparing Figure 7(D) and Figure 12(A), one can notice that the calculated values in the higher resolution case are relatively larger than those in the lower resolution case. The reason is that the lower resolution case tends to average out or filter out more small-scale variations and details.

\begin{figure}
 \centerline{\includegraphics[scale=0.64,clip=true,trim=0 0 0 0]{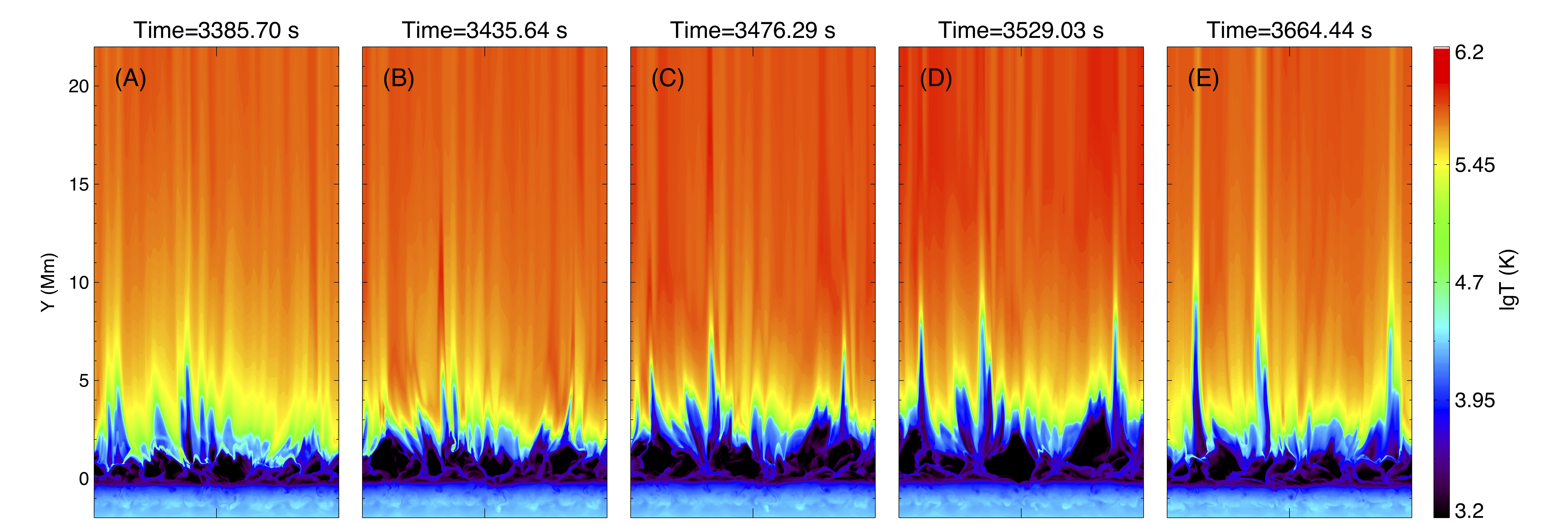}}
 \centerline{\includegraphics[scale=0.64,clip=true,trim=0 0 0 0]{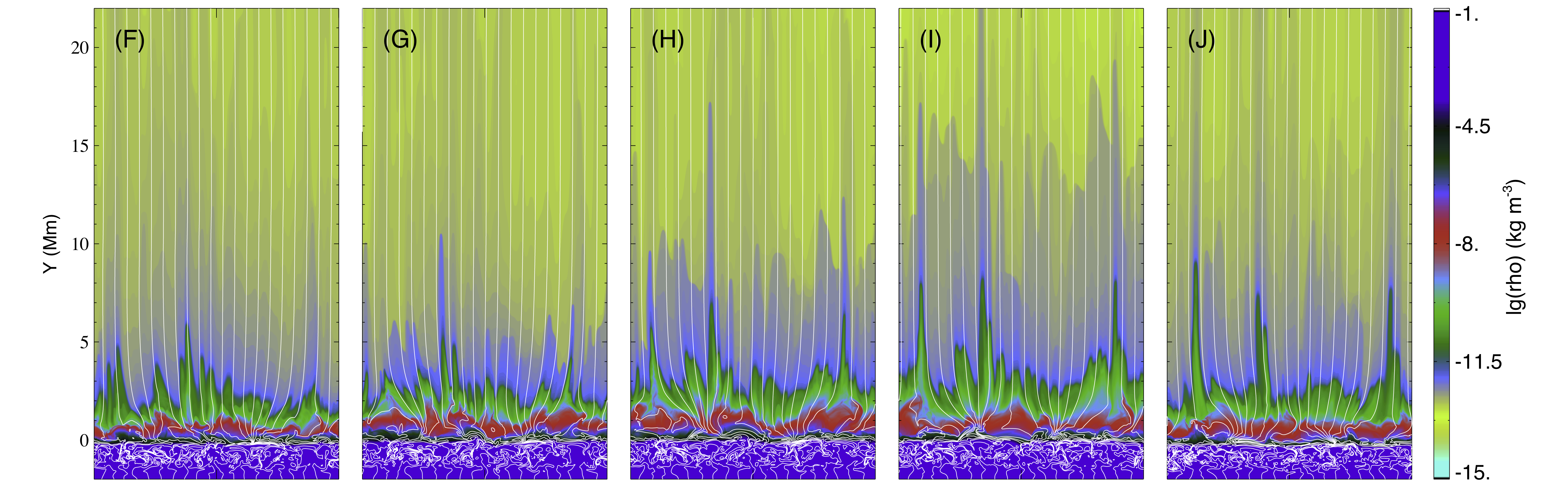}} 
 \centerline{\includegraphics[scale=0.64,clip=true,trim=0 0 0 0]{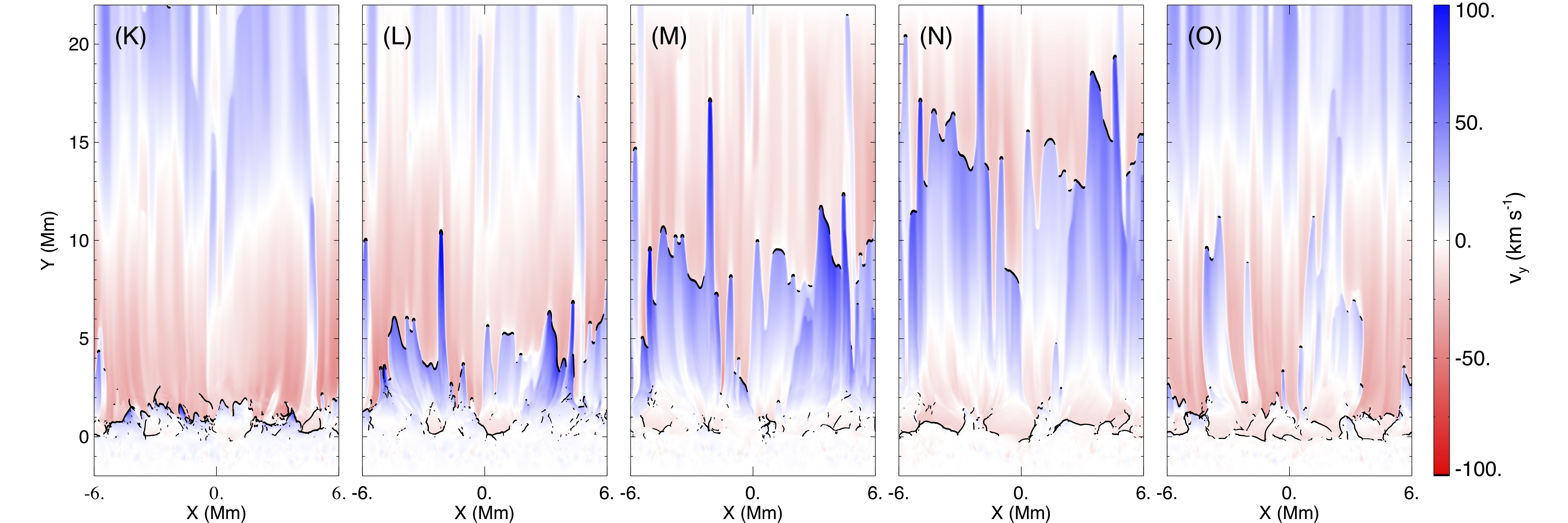}}                
\caption{\textbf{Evolution of spicules and the lower corona within a cycle of spicule formation in the higher resolution 2.5 D run. }All the initial and boundary conditions in this case are the same as those in the case with initial magnetic field $B_0=B_{0y}=5$\,G shown in Figure 3.  Distributions of logarithmic temperature (A-E), logarithmic density (F-J), and velocity in the \textit{Y}-direction (K-O) at five different times are presented. The white solid curves in (F-J) are for the magnetic field lines. The black contour lines in (K-O) represent the regions having large values of $-\nabla \cdot V$ (shock fronts). The animations of the corresponding temperature distribution (MovieS3) is available.}
\end{figure}

\begin{figure}
 \centerline{\includegraphics[scale=0.22,clip=true,trim=0 0 0 0]{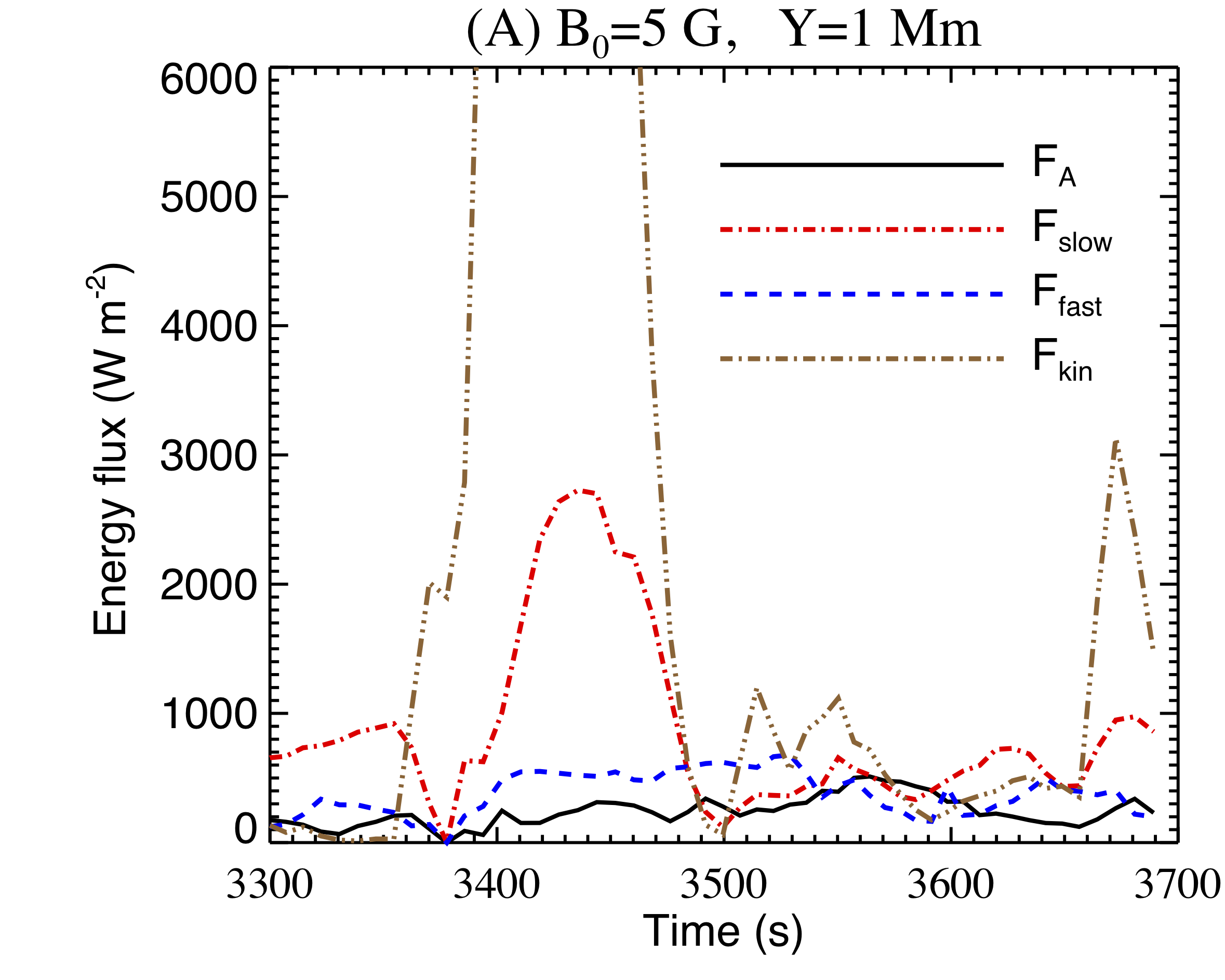}
                   \includegraphics[scale=0.22,clip=true,trim=0 0 0 0]{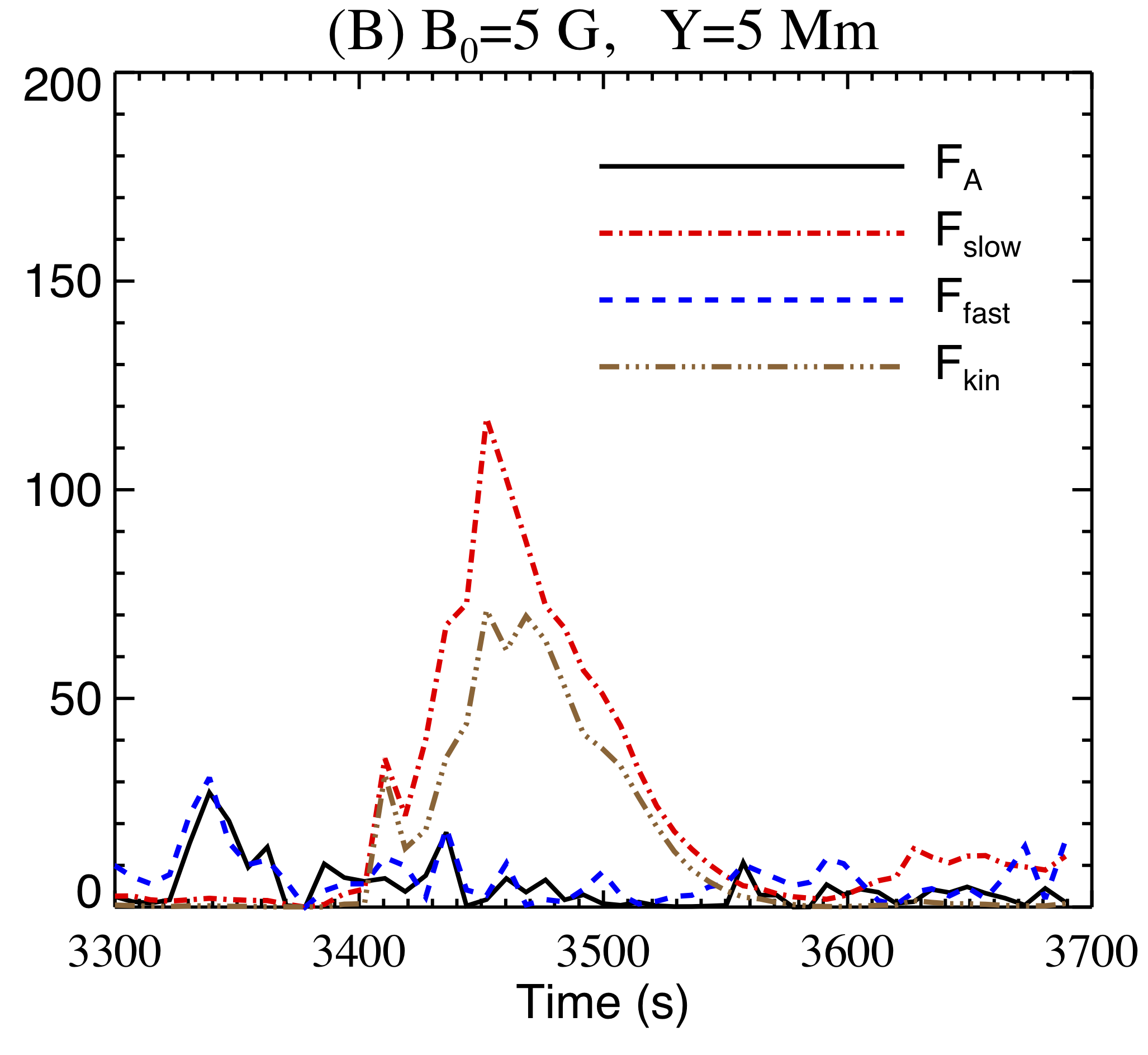}
                   \includegraphics[scale=0.22,clip=true,trim=0 0 0 0]{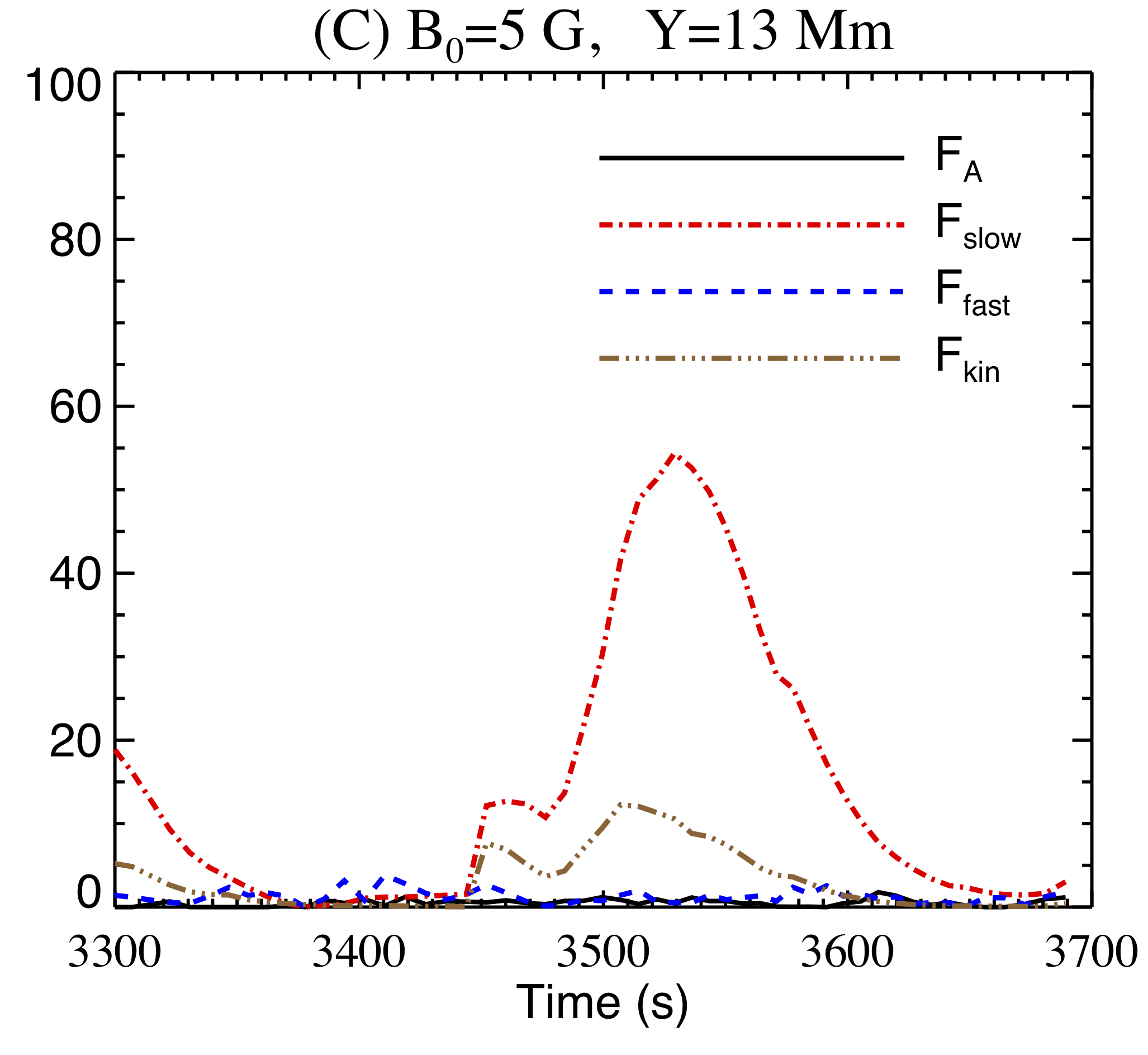}
                    \includegraphics[scale=0.22,clip=true,trim=0 0 0 0]{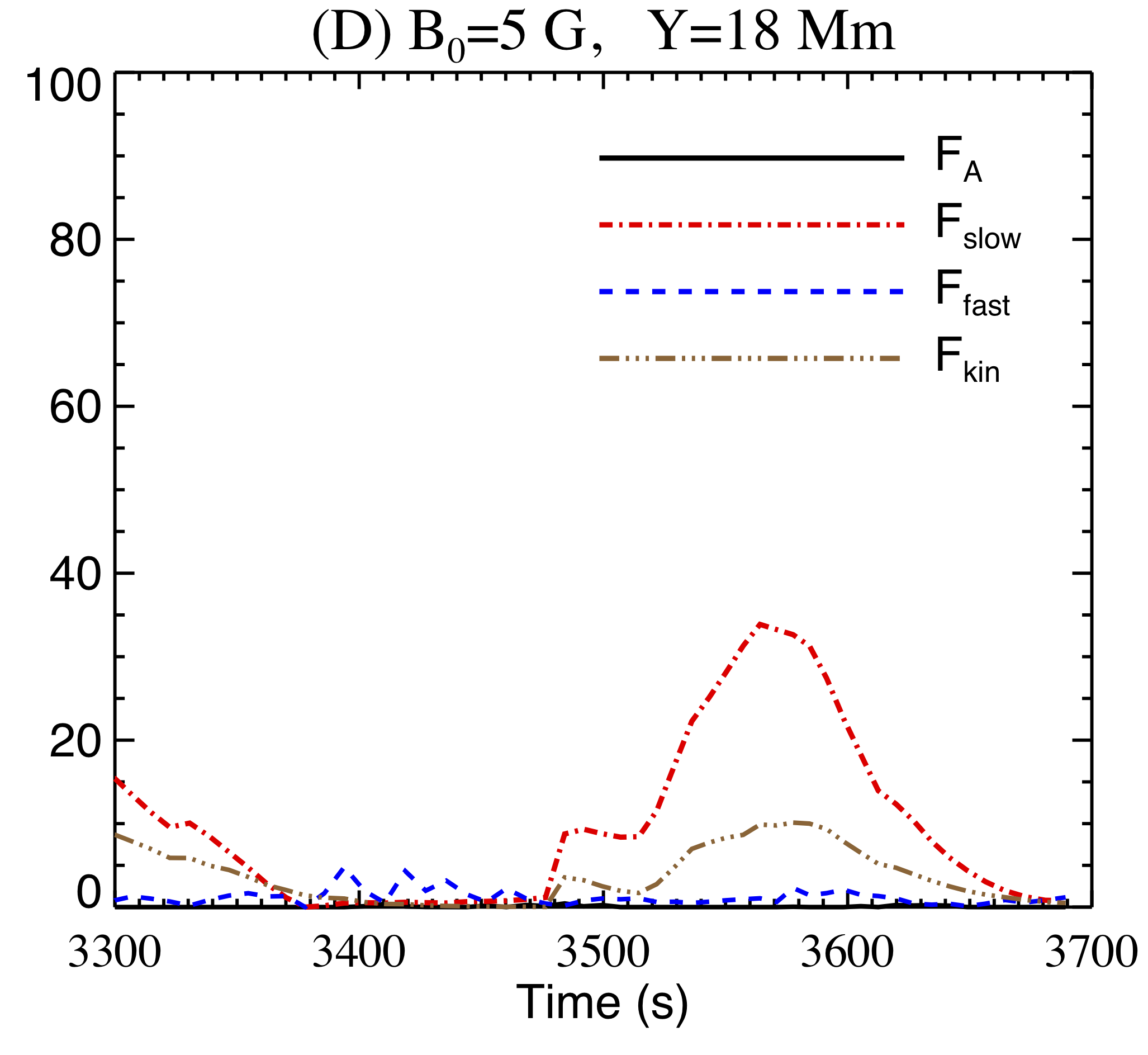}}
  \centerline{\includegraphics[scale=0.22,clip=true,trim=0 0 0 0]{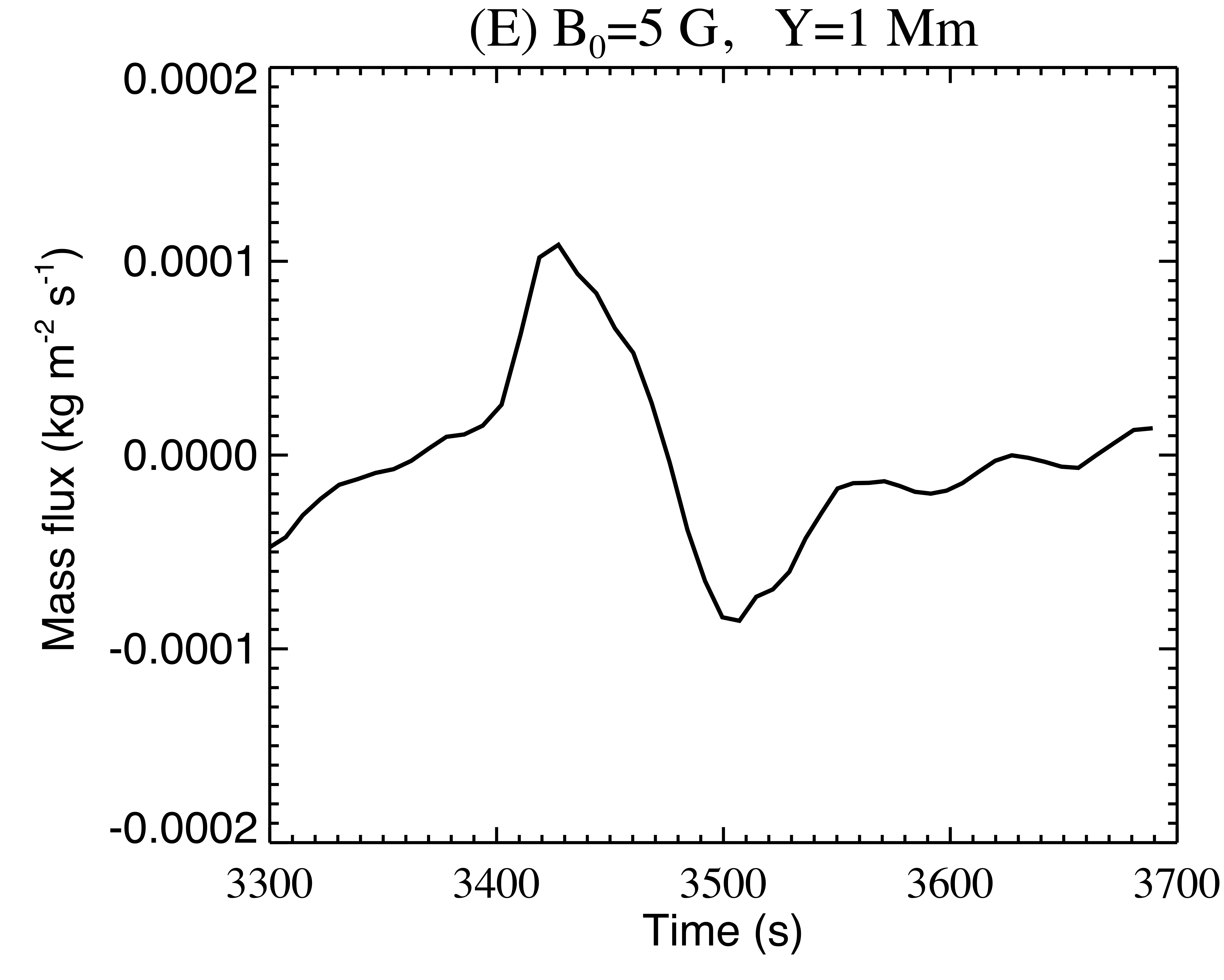}
                    \includegraphics[scale=0.22,clip=true,trim=0 0 0 0]{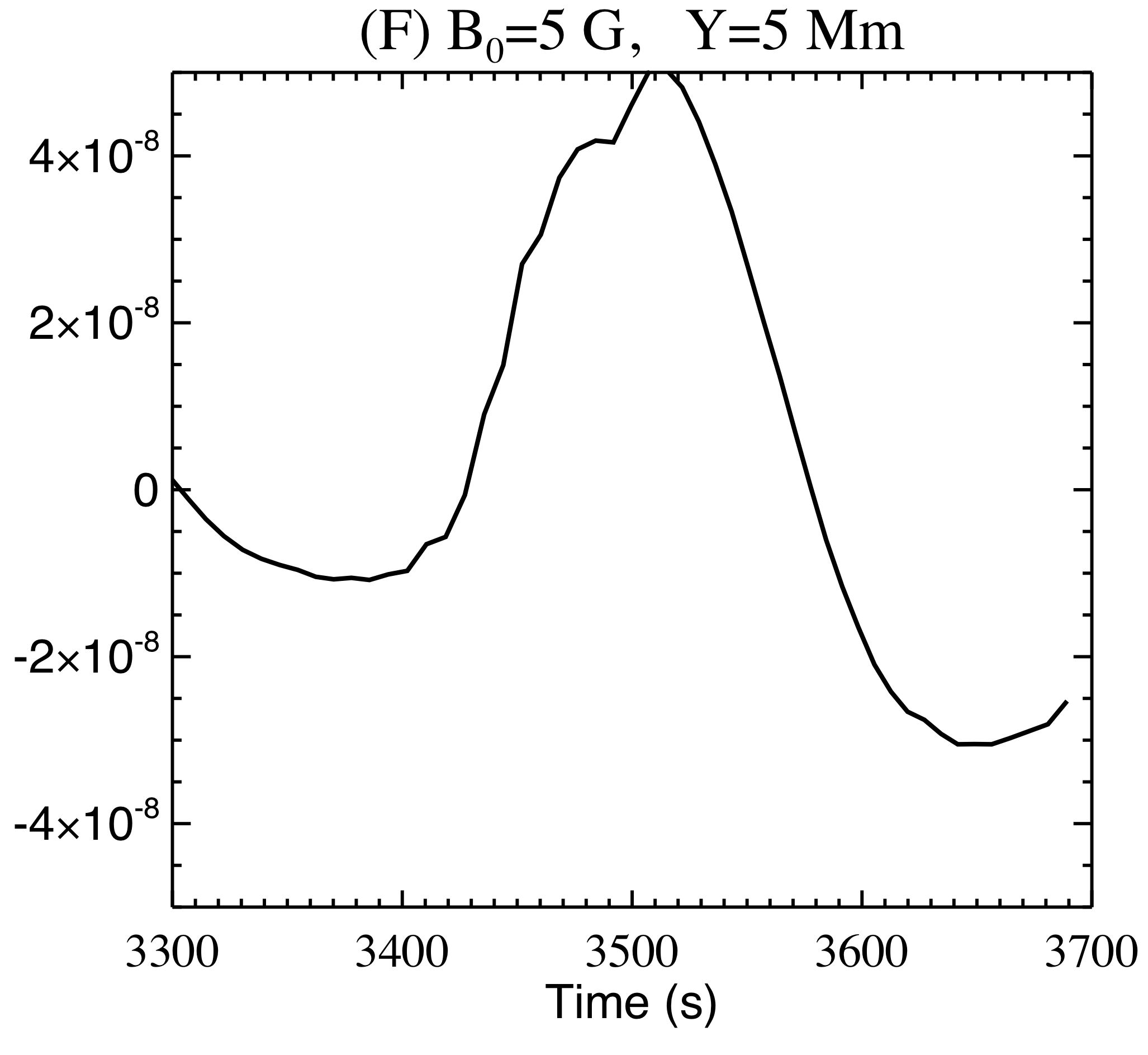}
                    \includegraphics[scale=0.22,clip=true,trim=0 0 0 0]{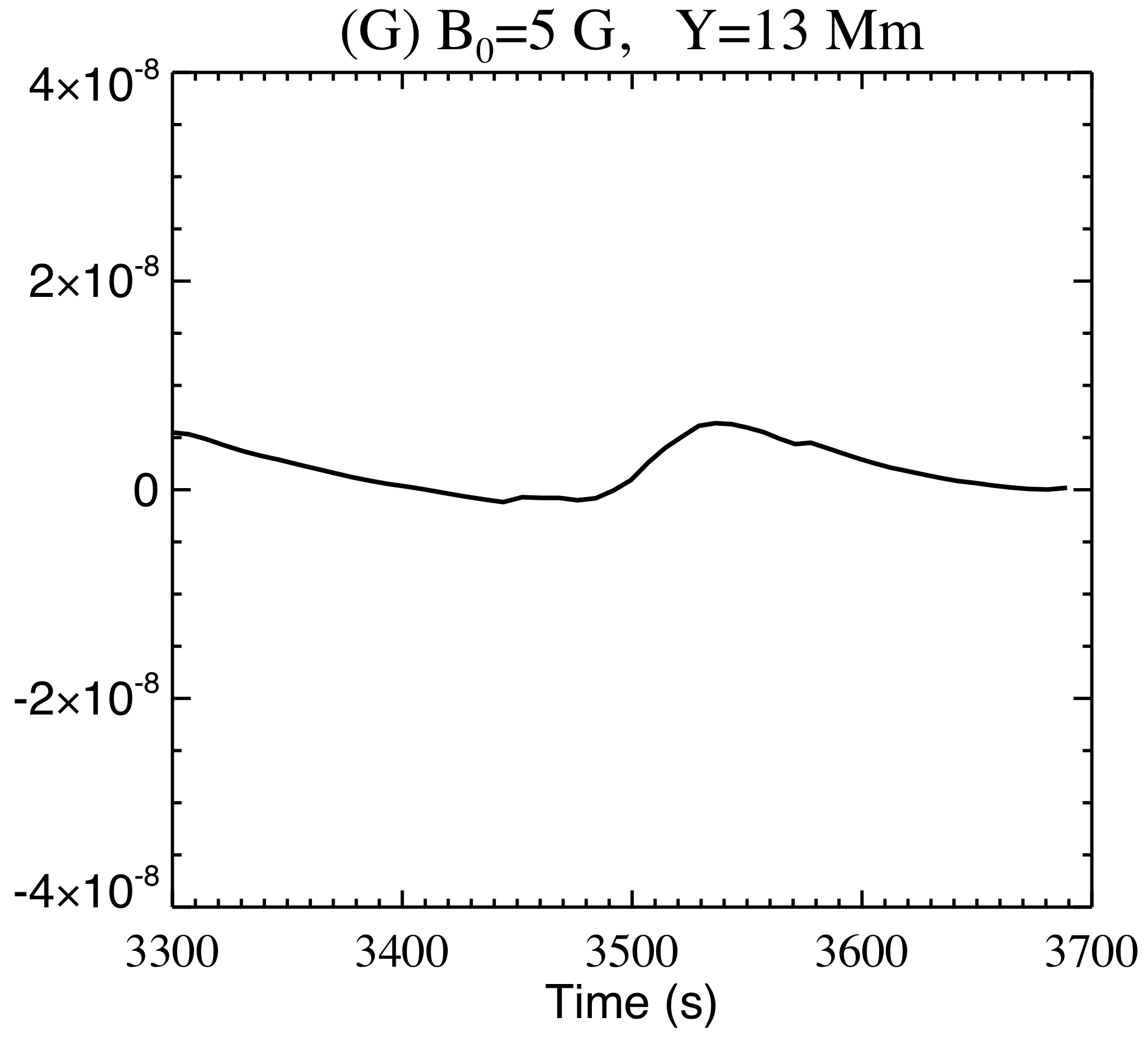}
                     \includegraphics[scale=0.22,clip=true,trim=0 0 0 0]{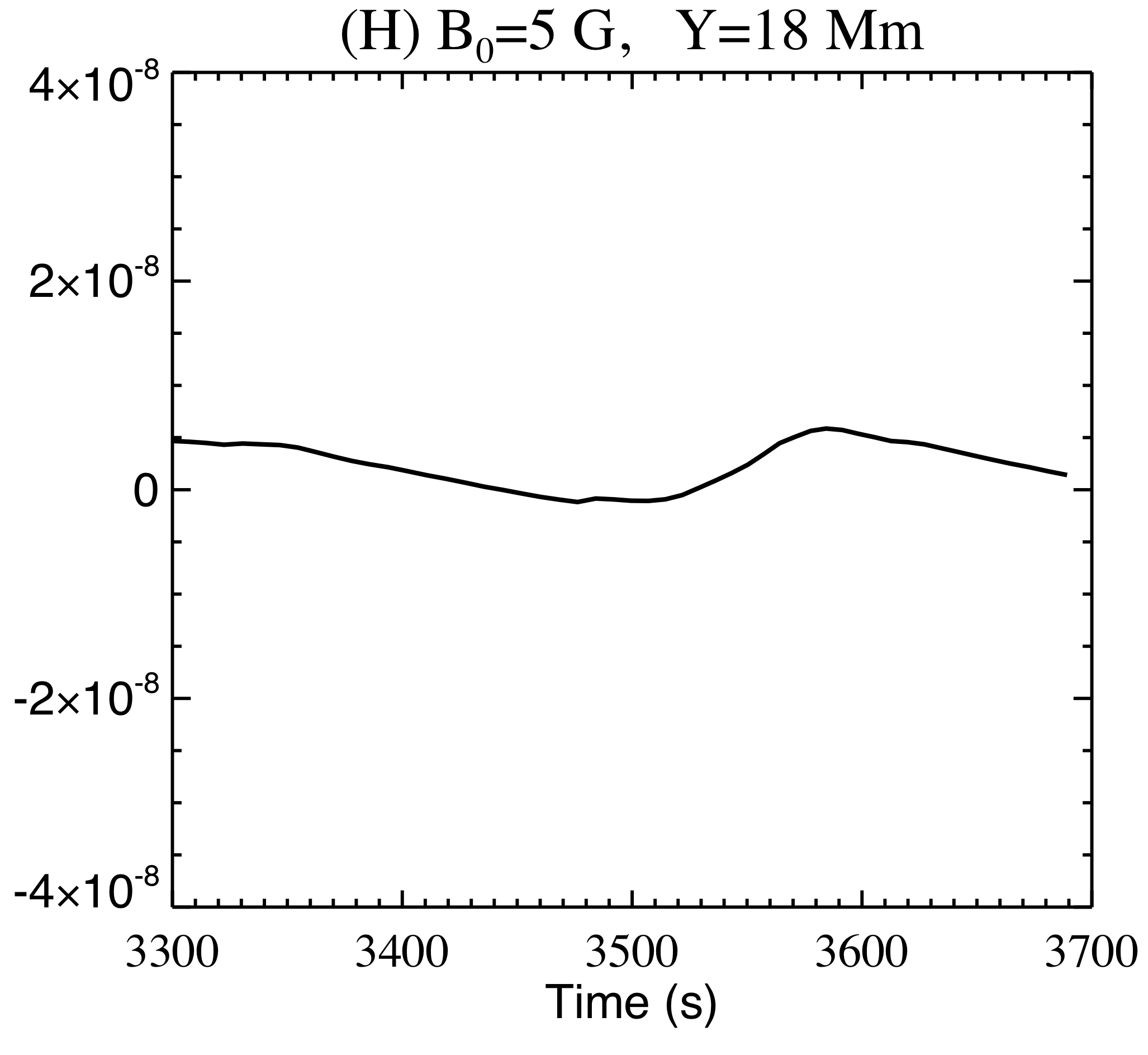}}             
\caption{\textbf{Upward energy fluxes and the mass fluxes from the chromosphere to the corona in the higher resolution case.} The energy fluxes carried by Alfv\'{e}n ($F_A$) waves, fast-mode waves ($F_{fast}$) and slow-mode waves ($F_{slow}$) (A-D) and the mass flux (E-H) varying with time at $Y=1$\,Mm, $Y=5$\,Mm, $Y=12$\,Mm and $Y=18$\,Mm are presented. }
\end{figure}

\begin{figure}
\centerline{\includegraphics[scale=0.42,clip=true,trim=0 0 0 0]{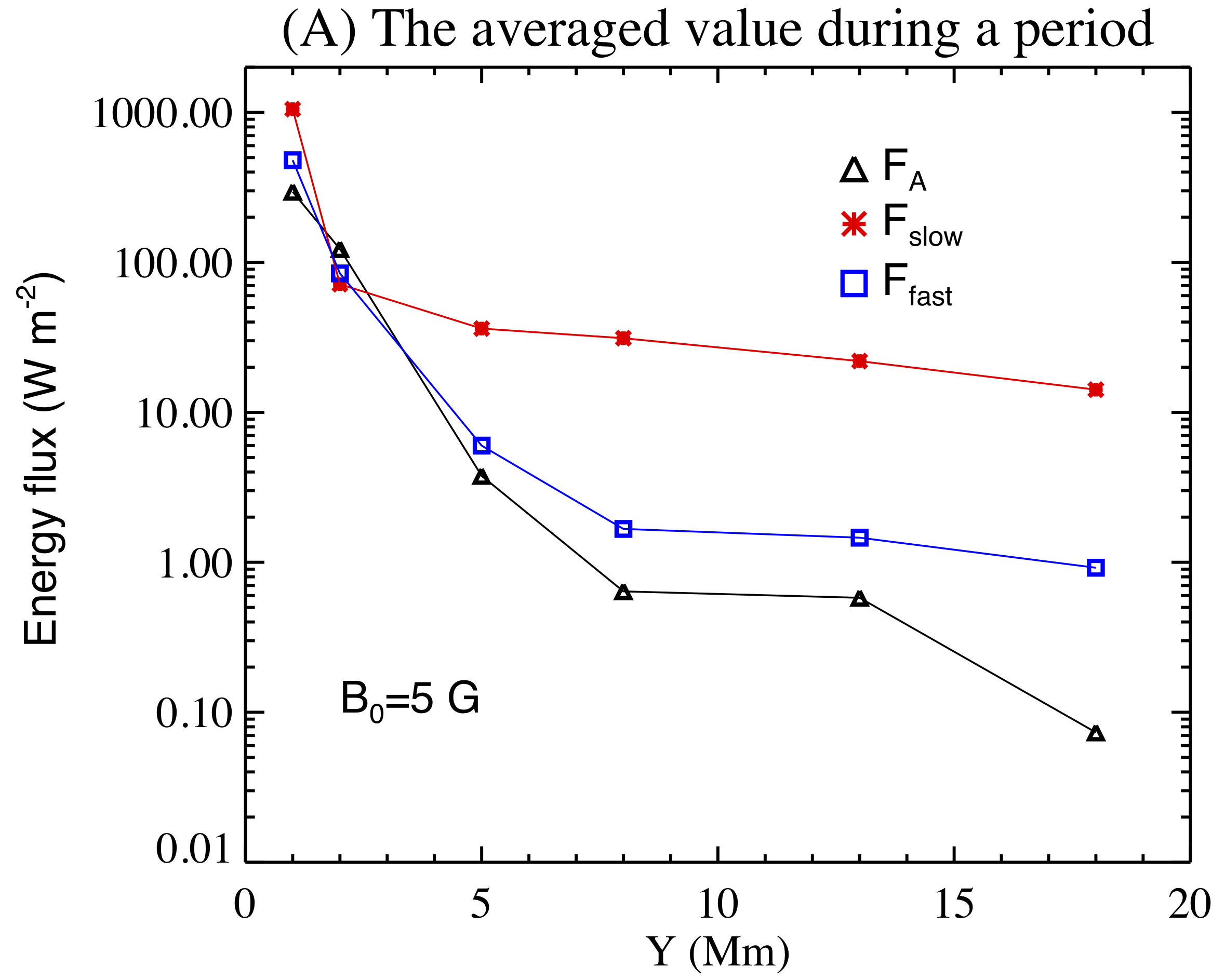}
                   \includegraphics[scale=0.42,clip=true,trim=0.5 0 0 0]{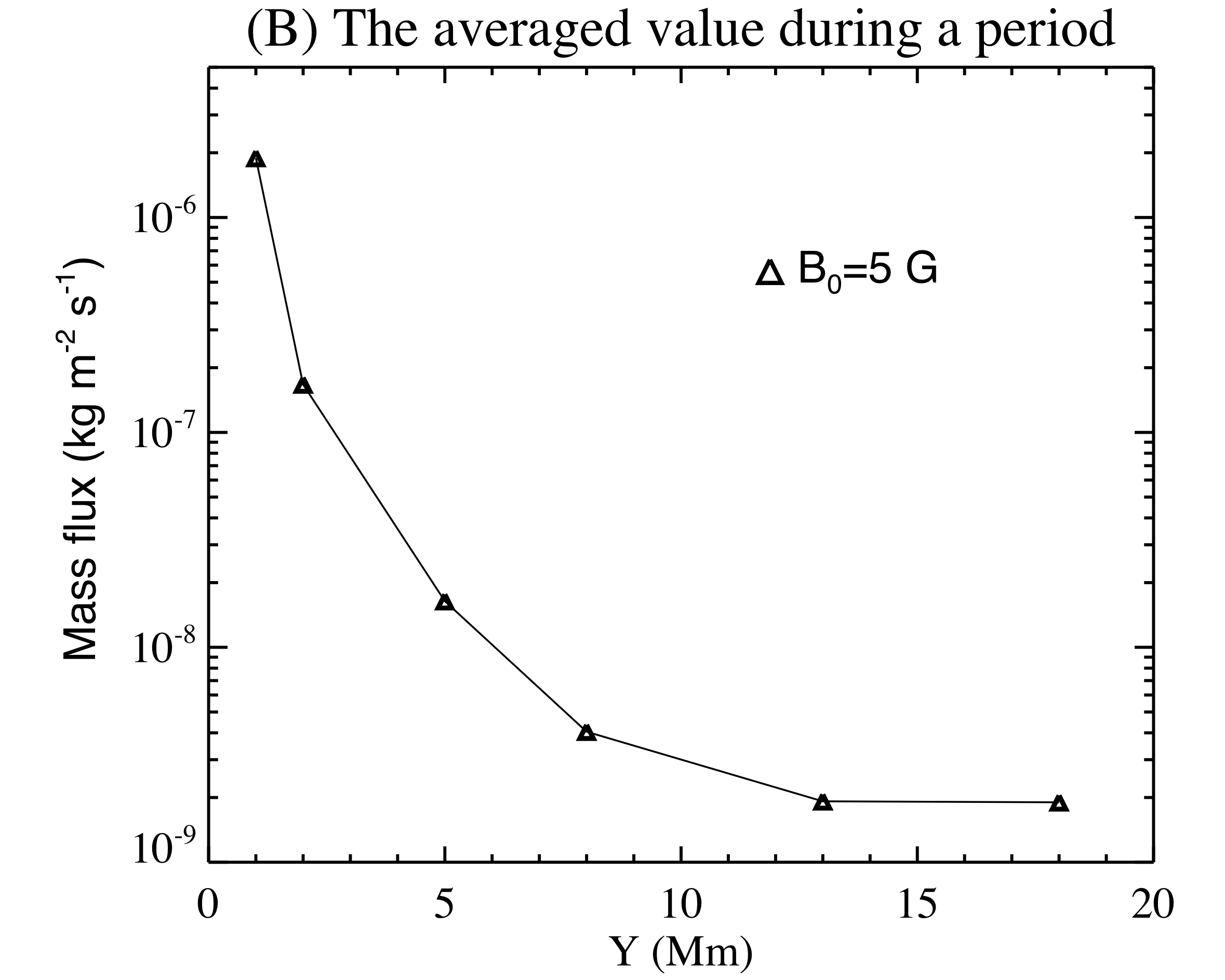}}
\caption{\textbf{Averaged upward energy fluxes (A) and mass flux (B) within a cycle of spicule formation at different altitudes in the higher resolution case are presented.} }
\end{figure}

\section{The results from 3D MHD simulations}

We have performed a full 3D simulation to complement this work. The C7 solar atmosphere model is also used to set the initial plasma parameters, and the density stratification occurs in the \textit{Z}-direction that is vertical to the solar surface. The initial background magnetic is assumed as $B_0=10$\,G in the \textit{Z}-direction. The simulation domain spans from $-8$\,Mm to $8$\,Mm in both the \textit{X} and \textit{Y} directions, and from $-3$\,Mm to $9$ Mm in the \textit{Z}-direction. The grid number is $96 \times 96 \times 256$. The grid size in the \textit{Z}-direction is $46.9$\,km, and it is $167$\,km in the \textit{X} and \textit{Y} directions. Therefore, the resolution in this 3D simulation is much lower than those in the 2.5 D simulations. However, the spicule-like structures are also formed in this 3D run. We also find that significant heating still appears in the corona above the cold spicule during its rising phase. These heatings might also be caused by slow-mode waves and shocks, which requires future higher-resolution 3D simulations to verify.

\clearpage

\newpage

\bibliography{corhheating-ver4}{}
\bibliographystyle{aasjournal}

\end{document}